\pdfoutput=1
\documentclass[11pt,twoside,a4paper,cmspaper,final,collab]{cms-tdr}
\def\svnVersion{f0cd141}\def\svnDate{2023/06/14}\def\cmsCernNoTag{CERN-EP-2022-254}\def\cmsCernDate{\today}\def\cmsMessage{Published in the Journal of High Energy Physics as \href{http://dx.doi.org/10.1007/JHEP06(2023)060}{\doi{10.1007/JHEP06(2023)060}.}}
\begin{document}\cmsNoteHeader{SUS-21-003}

\newcommand{\Iabs}{\ensuremath{I_\text{abs}}\xspace}
\newcommand{\Irel}{\ensuremath{I_\text{rel}}\xspace}
\newcommand{\dxy}{\ensuremath{d_{xy}}\xspace}
\newcommand{\dz}{\ensuremath{d_{z}}\xspace}
\newcommand{\mt}{\mT\xspace}
\newcommand{\njet}{\ensuremath{N_{\text{jet}}}\xspace}
\newcommand{\nbl}{\ensuremath{N(\PQb^{\text{loose}})}\xspace}
\newcommand{\ptl}{\ensuremath{\pt(\ell)}\xspace}
\newcommand{\etl}{\ensuremath{\eta(\ell)}\xspace}
\newcommand{\chgl}{\ensuremath{Q(\ell)}\xspace}
\newcommand{\ptisr}{\ensuremath{\pt(\mathrm{ISR})}\xspace}
\newcommand{\ptb}{\ensuremath{\pt(\PQb)}\xspace}
\newcommand{\drLB}{\ensuremath{\Delta R (\ell, \PQb)}\xspace}
\newcommand{\bdisc}{\ensuremath{D(\PQb)}\xspace}
\newcommand{\wjets}{\ensuremath{\PW}+jets\xspace}
\newcommand{\zinv}{\ensuremath{\PZ\to\PGn\PAGn}+jets\xspace}
\newcommand{\mmplane}{\ensuremath{({m}(\stp),~{m}(\lsp))}\xspace}
\newcommand{\lsp}{\PSGczDo\xspace}
\newcommand{\stp}{\ensuremath{\PSQt_{1}}\xspace}
\newcommand{\stpb}{\ensuremath{{\PASQt}_{1}}\xspace}
\newcommand{\dm}{\ensuremath{\Delta {m}}\xspace}
\newcommand{\htmiss}{\ensuremath{H^{\mathrm{miss}}_{\mathrm{T}}}\xspace}
\newcommand{\ttX}{\ensuremath{\ttbar\PX}\xspace}

\newlength\cmsTabSkip\setlength{\cmsTabSkip}{1ex}

\cmsNoteHeader{SUS-21-003}
\title{Search for top squarks in the four-body decay mode with single lepton final states in proton-proton collisions at \texorpdfstring{$\sqrt{s}=13\TeV$}{sqrt(s) = 13 TeV}}

\date{\today}

\abstract{A search for the pair production of the lightest supersymmetric partner of the top quark, the top squark ($\stp$), is presented. The search targets the four-body decay of the $\stp$, which is preferred when the mass difference between the top squark and the lightest supersymmetric particle is smaller than the mass of the $\PW$ boson. This decay mode consists of a bottom quark, two other fermions, and the lightest neutralino (\PSGczDo), which is assumed to be the lightest supersymmetric particle. The data correspond to an integrated luminosity of 138\fbinv of proton-proton collisions at a center-of-mass energy of 13\TeV collected by the CMS experiment at the CERN LHC. Events are selected using the presence of a high-momentum jet, an electron or muon with low transverse momentum, and a significant missing transverse momentum. The signal is selected based on a multivariate approach that is optimized for the difference between $m(\stp)$ and $m(\PSGczDo)$. The contribution from leading background processes is estimated from data. No significant excess is observed above the expectation from standard model processes. The results of this search exclude top squarks at 95\% confidence level for masses up to 480 and 700\GeV for $m(\stp) - m(\PSGczDo) = 10$ and 80\GeV, respectively.}

\hypersetup{
pdfauthor={CMS Collaboration},
pdftitle={Search for top squarks in the four-body decay mode with single lepton final states in proton-proton collisions at sqrt(s) = 13 TeV},
pdfsubject={CMS},
pdfkeywords={CMS,  SUSY}} 

\maketitle 

\section{Introduction}
\label{s:intro}

\begin{figure}[!htbp]
\centering
\includegraphics[width=0.60\textwidth]{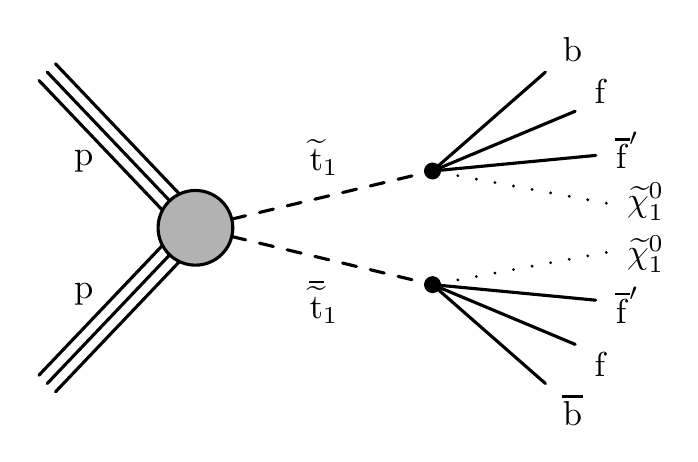}
\caption{Diagram of top squark pair production $\stp \stpb$ in $\Pp\Pp$ collisions, 
with a four-body decay of each top squark.}
\label{fig:model}
\end{figure}

Supersymmetry (SUSY)~\cite{Martin,SUSY0,SUSY1,SUSY2,SUSY3,SUSY4} predicts the
existence of a new symmetry that requires that, for each fermion (boson) in the 
standard model (SM), there is also a bosonic (fermionic) superpartner.
Searches for SUSY are among the important focal points of the physics program at the 
CERN LHC, since SUSY naturally solves the problem of quadratically divergent loop 
corrections to the mass of the Higgs boson~\cite{HgDisc2,HgDisc1,HgDisc3}.
If $R$ parity~\cite{Farrar:1978xj} is conserved, supersymmetric particles would
be produced in pairs, and their decay chains would end with the lightest 
supersymmetric particle (LSP), often considered to be the lightest neutralino \lsp.
Such an LSP, being neutral, weakly interacting, and massive, would have the required 
characteristics for a dark matter particle, and thus, would offer a solution to 
another shortcoming of the SM.
When the symmetry is broken, the scalar partners of an SM fermion acquire a mass 
different from the mass of the SM partner, with the mass splitting between scalar mass 
eigenstates being proportional to the mass of the SM fermion. 
Since the top quark is the heaviest fermion of the SM, the splitting between its 
chiral supersymmetric partners can be the largest among all supersymmetric quarks 
(squarks). Furthermore, the top Yukawa coupling can be the greatest among all fermions, 
which affects the masses of the squarks through the renormalization 
group equations. The lighter supersymmetric scalar partner of the top quark, the
top squark (\stp), could therefore be the lightest squark.
If SUSY is realized in nature, cosmological observations  
imply that for many models the lightest top squark should be almost degenerate 
with the LSP~\cite{Coannihilation}.
In this scenario, because the mass difference between the \stp and the \PSGczDo 
is smaller than the mass of the \PW boson, the two- and three-body decays of 
the \stp are kinematically forbidden, while the two-body decay to $\cPqc\PSGczDo$
can be suppressed depending on the parameters of the model.
This motivates the search for the four-body decay 
$\stp \to \PQb \mathrm{f} \overline{\mathrm{f}}^{\,\prime} \PSGczDo$, where 
\PQb stands for the bottom quark, and the fermions ${\mathrm{f}}$ and 
$\overline{\mathrm{f}}^{\,\prime}$ can be either quarks or leptons. 
Throughout this paper, charge conjugation is assumed.
Figure~\ref{fig:model} represents a simplified model~\cite{sms,ArkaniHamed:2007fw,Alwall:2008ag,Alwall:2008va,Alves:2011wf,Chatrchyan:2013sza} of the production of $\stp \stpb$ in proton-proton 
($\Pp\Pp$) collisions, where each \stp and \stpb undergoes a four-body decay.

In this paper, the previous 2016 result of CMS at $\sqrt{s} = 13 \TeV$~\cite{st4bd} is combined with 
data recorded in 2017 and 2018. The results of 2016 are directly taken from Ref.~\cite{st4bd}, 
except the integrated luminosity and its uncertainty, which are updated to their latest 
values~\cite{CMS-LUM-17-003}. The total integrated luminosity for the combined 2016--2018 
analysis is 138\fbinv.

In the present search a final state is considered, where the fermions ${\mathrm{f}}$ and 
$\overline{\mathrm{f}}^{\,\prime}$ represent a charged lepton and its neutrino for the 
decay products of one \stp, and two quarks for the other top squark. A 100\% branching 
fraction is assumed for the four-body decay when interpreting the results~\cite{sms}. 
The considered final states contain at least one jet, a large missing transverse momentum, 
and exactly one charged lepton, which can be either an electron or a muon. The choice of 
final states where one top squark decays into a lepton is motivated by the decrease of the 
contributions from the multijet background in this mode, while increasing the selection 
efficiency with the other top squark decaying hadronically.
The selected jet, attributed to initial-state radiation (ISR) of a parton, is required to 
have high transverse momentum (\pt). Both neutralinos and the neutrino escape undetected, 
leaving high missing transverse momentum. Electrons and muons can be efficiently reconstructed 
and identified with \pt as low as 5.0 and 3.5\GeV, respectively. 
The signal selection is based on a multivariate analysis, followed by a counting experiment. 
This approach takes advantage of the different correlations between the discriminating variables 
for signal and background, and is adapted for different $\dm = m(\stp) - m(\lsp)$ kinematic 
regions, thus enhancing the reach of the search across the \mmplane space. The main contributions 
to the background events are \wjets, \ttbar, and {\PZ}+jets processes, and 
are predicted from data. 
A search in the single-lepton final state for the four-body decays of the \stp has been performed 
by the ATLAS Collaboration at $\sqrt{s} = 13 \TeV$~\cite{atlasRun2}, and a comparison of its 
results to the present search is provided in this paper. 
Tabulated results are provided in the HEPData record for this search~\cite{hepdata}.

\section{The CMS detector}
\label{s:detector}

The central feature of the CMS apparatus~\cite{CMS:2008xjf} is a superconducting 
solenoid of 6\unit{m} internal diameter, providing a magnetic field of 3.8\unit{T}.
Within the solenoid volume are a silicon pixel and strip tracker, a lead
tungstate crystal electromagnetic calorimeter (ECAL), and a brass and
scintillator hadron calorimeter (HCAL), each composed of a barrel and two
endcap sections. Forward calorimeters extend the pseudorapidity coverage
provided by the barrel and endcap detectors. The silicon tracker measures
charged particles within the pseudorapidity range $\abs{\eta}<2.5$. Muons
are detected in gas-ionization chambers embedded in the steel flux-return
yoke outside the solenoid.
It is a multipurpose, nearly hermetic detector, designed to trigger 
on~\cite{CMS:2020cmk,CMS:2016ngn} and identify electrons, muons, 
photons, and hadrons~\cite{CMS:2020uim,CMS:2018rym,CMS:2014pgm}.

\section{Data and simulated samples}
\label{s:samples}

The search described in this paper is performed using data from $\Pp\Pp$
collisions recorded in 2017--2018 by the CMS experiment at the LHC at a 
center-of-mass energy of 13\TeV. These results are statistically combined with 
the result from the search using the data of 2016, corresponding to an 
integrated luminosity of 36.3\fbinv that was updated for this paper. The 2017 
and 2018 data samples correspond to an integrated luminosity of 41.5\fbinv and 
59.8\fbinv, respectively, bringing the total integrated luminosity for the 
combined 2016--2018 analysis to 138\fbinv.

Signal and background processes are simulated using several Monte Carlo (MC) 
event generators. The SM background MC samples are used to estimate the relation 
between the control and signal regions for the main background processes, to 
validate the background estimation methods based on control samples in data, and 
finally, to predict the contributions from other processes.
The \wjets, \zinv, and multijet processes are 
generated at leading order (LO) by \MGvATNLO~2.4.2~\cite{Alwall:2014hca}. 
The \ttbar process is generated at next-to-leading order (NLO) by \MGvATNLO.
The \POWHEG~v2.0~\cite{Nason:2004rx, Frixione:2007vw, Alioli:2010xd, Alioli:2009je, 
Re:2010bp, Melia:2011tj, Nason:2013ydw, Hartanto:2015uka} generators are used for 
the NLO simulations of single top and associated $\cPqt \PW$ production.
Diboson events are simulated at NLO with 
\MGvATNLO and \POWHEG. The LO (NLO) NNPDF3.1LO (NNPDF3.1NLO)~\cite{Ball:2017nwa} 
parton distribution functions are used consistently with the order of the matrix element calculation in the 
generated events. Additional backgrounds such as \ttbar produced in 
association with a \PZ boson, \PW boson, or photon, referred to as \ttX, are 
generated with \MGvATNLO at NLO~\cite{Frederix:2012ps}.
Hadronization and showering of events in all generated samples are
simulated using \PYTHIA~8.230~\cite{Sjostrand:2014zea} with 
the CP5 tune~\cite{CMS:2019csb} for the underlying event.
All SM MC events are passed through a full simulation of the CMS apparatus, where 
the response of the detector is modeled using the \GEANTfour~\cite{GEANT4} 
software. Generated events are processed using the same version of the CMS event 
reconstruction software used as for data. 
Additional $\Pp\Pp$ collisions in the same or nearby beam crossings (pileup) 
are simulated and overlaid
on the main $\Pp\Pp$ interaction in the MC samples, with distributions that 
reproduce the conditions observed year-to-year in data.
For the pair production of top squarks ($\stp\stpb$), simulated samples are produced for 
$250 < m(\stp) < 800\GeV$ in steps of 25\GeV, and $10 < \dm < 80\GeV$ 
in steps of 10\GeV. The cross section for $\stp\stpb$ production, calculated using 
\PROSPINO v.2~\cite{prospino,Borschensky:2014cia,xs1,xs2,xs3,xs4,xs5}, is computed at 
next-to-next-to-leading-order (NNLO) accuracy, and includes next-to-next-to-leading-logarithmic 
(NNLL) corrections~\cite{Beenakker:2016lwe,Beenakker:2016gmf}. This cross 
section varies between approximately 25 and 0.03\unit{pb} as m(\stp) goes from 
250 to 800\GeV. The generation of signal events with up to two additional jets, 
which can originate from  ISR, is performed with \MGvATNLO and then interfaced 
with \PYTHIA for the decay hadronization and showering. The modeling of the 
detector response for the signal is done with the CMS fast simulation 
program~\cite{Abdullin:2011zz,Giammanco:2014bza}.
The simulated samples of all background processes are used to check the modeling 
of variables at the preselection level (see Section \ref{s:presel}). For the 
final results, the yields of \wjets, \ttbar, \zinv and multijet backgrounds, as 
well as the associated systematic uncertainties, are predicted using data-driven
techniques (see Sections \ref{s:bkg} and \ref{s:sys}).

Both signal and background simulated samples are corrected to account for discrepancies 
from data.
Control regions (CRs) in data are used to measure the reconstruction efficiencies of
leptons and jets produced by the hadronization of \PQb quarks, "\PQb jets",
as well as the \PQb jet misidentification probabilities for light-quark and gluon jets. 
The corrections are applied as a function of the 
\pt and $\eta$ of the objects. Fast simulation signal samples are additionally 
corrected to take into account any potential difference with respect to the \GEANTfour 
modeling. The latter corrections translate into efficiencies applied to \PQb jets, 
leptons, and the modeling of the missing transverse momentum.
The simulations of \wjets, \ttbar, and signal processes are corrected for the effect 
of ISR. The modeling of ISR for these processes is checked in data-based control 
samples that are highly enriched in \ttbar or \wjets events. The simulation of \ttbar 
events is tested by comparing the jet multiplicity observed in a control sample with 
the simulation, and the \ttbar and signal samples are reweighted based on this comparison. 
The simulation of \wjets events is corrected based on the distribution of the sum of the 
magnitudes of the lepton \Pell transverse momentum \ptl and the missing transverse 
momentum in a control sample.

\section{Event reconstruction and object selection}
\label{s:objects}

Data and simulated events are reconstructed using the CMS particle-flow (PF) 
algorithm~\cite{CMS-PRF-14-001}, which matches the information from all
CMS subdetectors in order to describe the event in terms of global physics objects.
These objects are denoted as PF candidates and are classified into mutually 
exclusive categories: electrons, muons, photons, and charged and neutral hadrons.
The primary vertex (PV) is taken to be the vertex corresponding to the hardest 
scattering in the event, evaluated using tracking information alone, 
as described in Section 9.4.1 of Ref.~\cite{CMS-TDR-15-02}. 

Jets are reconstructed by applying the anti-\kt clustering algorithm~\cite{Cacciari:2008gp,Cacciari:2011ma}
to PF candidates with a distance parameter of 0.4. The pileup contribution to the jet 
momentum is partially taken into account by excluding the charged hadrons 
originating from vertices other than the PV from the jet-clustering algorithm. 
To account for pileup contributions from neutral particles and any inhomogeneity in the 
detector response, the jet \pt is further calibrated as described in 
Ref.~\cite{Khachatryan:2016kdb}. Jets are required to satisfy $\pt > 30\GeV$, 
and $\abs{\eta}<2.4$. The tagging of \PQb jets (\PQb tagging) is performed with the DeepCSV 
algorithm~\cite{Sirunyan_deepcsv} that uses information from the secondary vertex and 
is based on a deep neural network. The \PQb tagging discriminant is used to tag 
jets as \PQb jets based on a set of working points (loose, medium, tight) and to
define further event variables based on the discriminant value or the jet with 
highest discriminant value. 
The \PQb jet identification working points are defined as the selection values
in the discriminator distribution at which the probability of misidentifying a 
light-flavor jet as a \PQb jet is 10\%, 1\% and 0.1\%, respectively, for the 
loose, medium and tight working points~\cite{CMS-DP-2018-058}.

The missing transverse momentum vector, \ptvecmiss, is computed as the negative 
vector $\vec{\pt}$ sum of all PF candidates in the event, and its magnitude is \ptmiss. 
The calibrations associated with the jet energy estimations are propagated
to the \ptvecmiss~\cite{CMS:2019ctu}.

Electron candidates are reconstructed from energy deposits in the ECAL and matched 
charged particle tracks in the inner tracker obtained using the Gaussian sum 
filter algorithm~\cite{CMS:2020uim}. To reduce the number of misidentified electrons, 
additional constraints on the shape of the electromagnetic shower in the ECAL, 
the quality of the match between the trajectory of the track, and the 
ECAL energy deposit around the electron, and the relative HCAL deposit in the electron direction are
applied. Electrons are required to have \pt above 5\GeV and 
$\abs{\eta}<2.5$, with a veto on electron candidates in the ECAL gap region ($1.4442<\abs{\eta}<1.5660$). 
They are identified with requirements on the observables that
describe the matching of the measurements in the tracker and the ECAL, the 
description of energy clusters in the ECAL, and the amount of bremsstrahlung 
radiation emitted during the propagation through the detector. A loose working 
point of this algorithm is required for electrons to be selected, which has an 
average efficiency of 90\%.

Muon candidates are reconstructed by combining the information from the silicon 
tracking systems and the muon spectrometer in a global fit~\cite{CMS:2018rym}
that assigns a quality to the matching between the tracker and muon systems 
and imposes minimal requirements on the track to reduce the misidentification 
of muons. The medium working point of this algorithm is required for muons to be
selected, which ensures an efficiency above 98\%. Muons are required to pass the
selection requirements of $\pt > 3.5\GeV$ and $\abs{\eta}<2.4$.

To select electrons or muons originating from the PV, the point of 
closest approach of the associated track with respect to the PV is required to have a transverse 
distance $\abs{\dxy}<0.02\unit{cm}$, and a longitudinal distance $\abs{\dz}<0.1\unit{cm}$.
A lepton is defined as being nonprompt either when it does not originate from the PV, or 
when a jet is misidentified as a lepton.
Background processes with nonprompt leptons are one of the main contributions to
the SM background in the signal regions. 
In this analysis, nonprompt leptons mostly arise from heavy-quark decays in jets produced in association with a
\zinv decay, from multijet production, or from \wjets and \ttbar 
events where the prompt lepton was not reconstructed and a different one was accepted. 
In order to suppress these types of processes, a requirement on the lepton isolation 
is applied, which uses a combination of an absolute and a relative isolation variable.
The absolute isolation variable \Iabs of the lepton is defined as 
the scalar sum of the \pt of PF candidates within a cone size of 
$R \equiv \sqrt{\smash[b]{(\Delta\phi)^2+(\Delta\eta)^2}} = 0.3$,
where $\phi$ is the azimuthal angle, around the 
lepton candidate, which is excluded from the sum, as are charged PF candidates not associated with the PV.
The contributions from neutral particles originating from pileup 
are estimated according to the method described in Refs.~\cite{CMS:2016krz,Rehermann:2010vq}, 
and are subtracted from \Iabs. The ratio of the lepton \Iabs to the lepton \pt
is defined as the lepton relative isolation \Irel. 
A uniform lepton selection efficiency as a function of \pt is achieved by requiring
leptons to have $\Iabs<5\GeV$ for $\ptl<25\GeV$ and $\Irel<0.2$ for
$\ptl\ge25\GeV$.

\section{Event selection}
\label{s:eventsel}

The data events collected by the trigger system are required to have both
\ptmiss and \htmiss above 120\GeV, where \htmiss is the magnitude of the missing transverse 
momentum calculated only from jets. In order to maintain the performance of the 
online selection with increased luminosity from the late runs of 2017 onward, 
the condition $\HT>60\GeV$ is also required, where \HT is defined as the 
scalar \pt sum of all jets in the event. 
The efficiency of the combined \ptmiss and \htmiss trigger is measured using an independent 
event sample with single-electron triggers and \pt thresholds of 35 and 32\GeV 
for the 2017 and 2018 data-taking periods, respectively. 

The offline event selection is a two-step process. First, a preselection is applied 
to reduce the contribution of the main background processes (Section~\ref{s:presel}) 
by selecting a single charged lepton, large \ptmiss, and jets. Then, boosted decision 
trees (BDTs)~\cite{RefBDT,Hocker:2007ht} are trained and used to define the signal 
selection (Section~\ref{s:sel}). The preselection is constructed to be as inclusive 
as possible in order to maintain a high signal efficiency for all \dm values, leaving 
the main selection to the BDT.

\subsection{Preselection}
\label{s:presel}

The value of the preselection \ptmiss threshold is set close to the beginning of the 
maximum efficiency plateau of the combined \ptmiss and \htmiss trigger, while optimizing 
the separation between signal and background performed by the BDTs. Events with 
$\ptmiss>280\GeV$ are selected, favoring the signal where two \lsp's escape 
detection and where the \ptmiss is therefore larger than for SM 
processes. For these events, the trigger efficiency is above 98\% for both years. 
To account for the small inefficiency, simulated samples are reweighted as a function
of \ptmiss to match the efficiency of the triggers in data.

To suppress the contribution of SM processes, additional requirements are imposed 
on the selected events. In particular, to reduce the \wjets background, $\HT>200\GeV$
is required.
To select the single-lepton topology, it is demanded exactly one identified electron or 
muon in the event, along with at least one jet. This selection reduces the contribution 
from the dilepton topology of \ttbar events. 
To further improve the selection of signal over SM background events, at least one jet must have
$\pt>110\GeV$. 
These requirements are geared towards signal events in which the $\stp \stpb$ system 
recoils against a high-momentum ISR jet, Lorentz boosting the \lsp and increasing 
\ptmiss. The ISR jet will often be the highest momentum (leading) jet in these events, 
and the leading-jet \pt threshold value is optimized in the same manner as for \ptmiss.
Lastly, in events with at least two jets, the azimuthal angle between the directions of
the leading and second-highest-\pt (subleading) jets must be smaller than 2.5 radians,
suppressing the SM multijet background.

\begin{figure*}[htbp]
\centering
\includegraphics[width=0.45\textwidth]{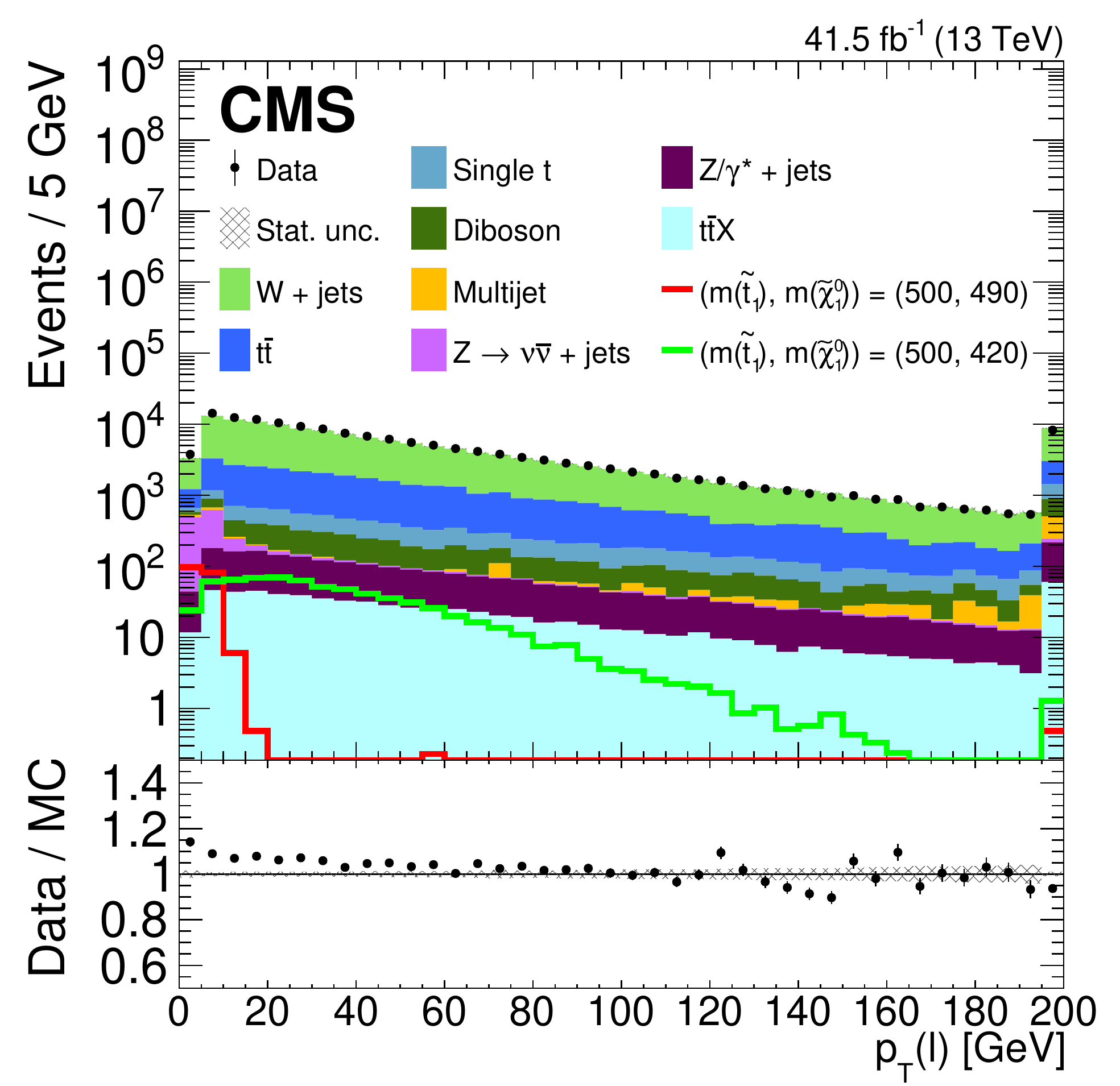} \hfil
\includegraphics[width=0.45\textwidth]{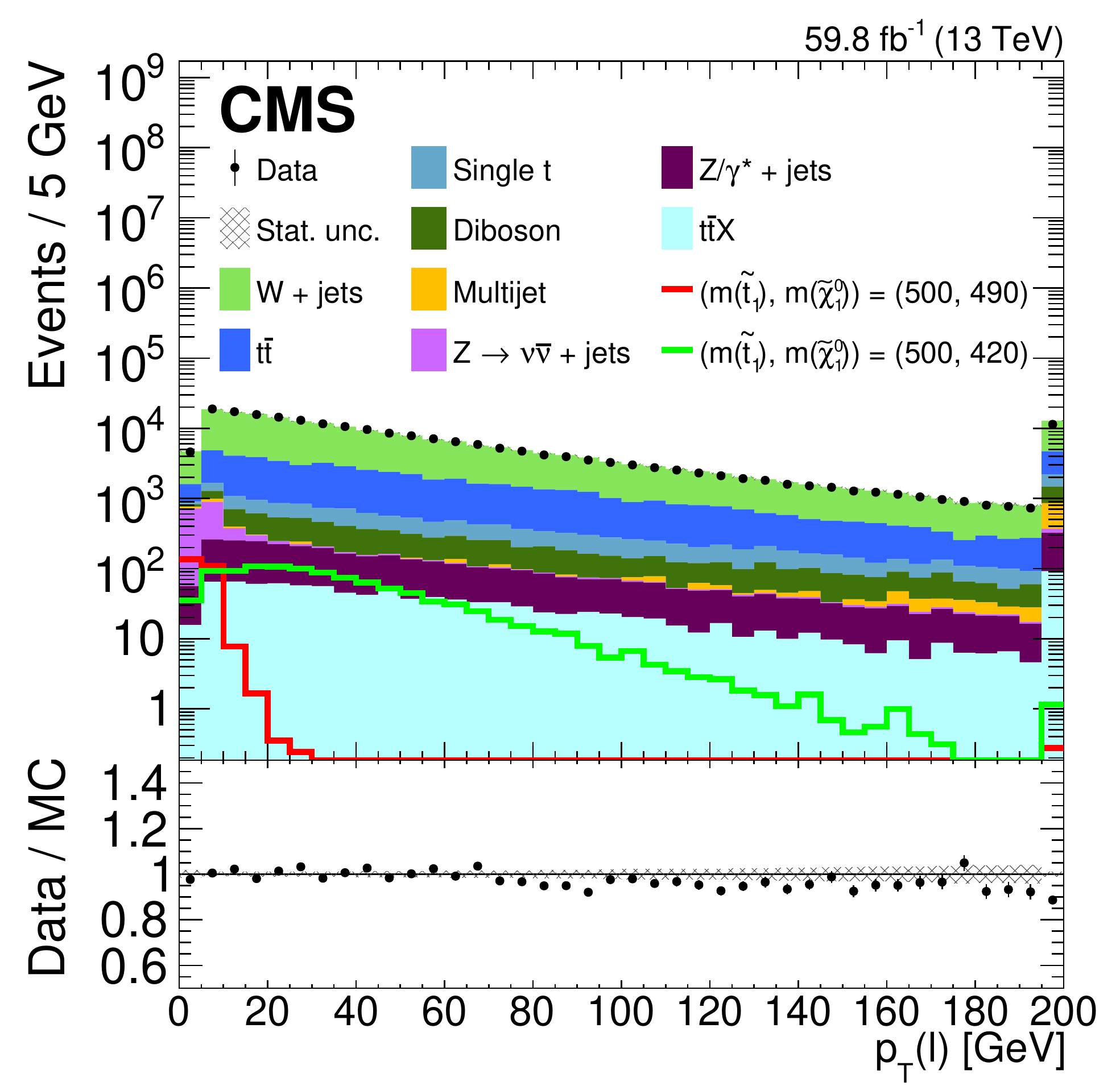} \\
\includegraphics[width=0.45\textwidth]{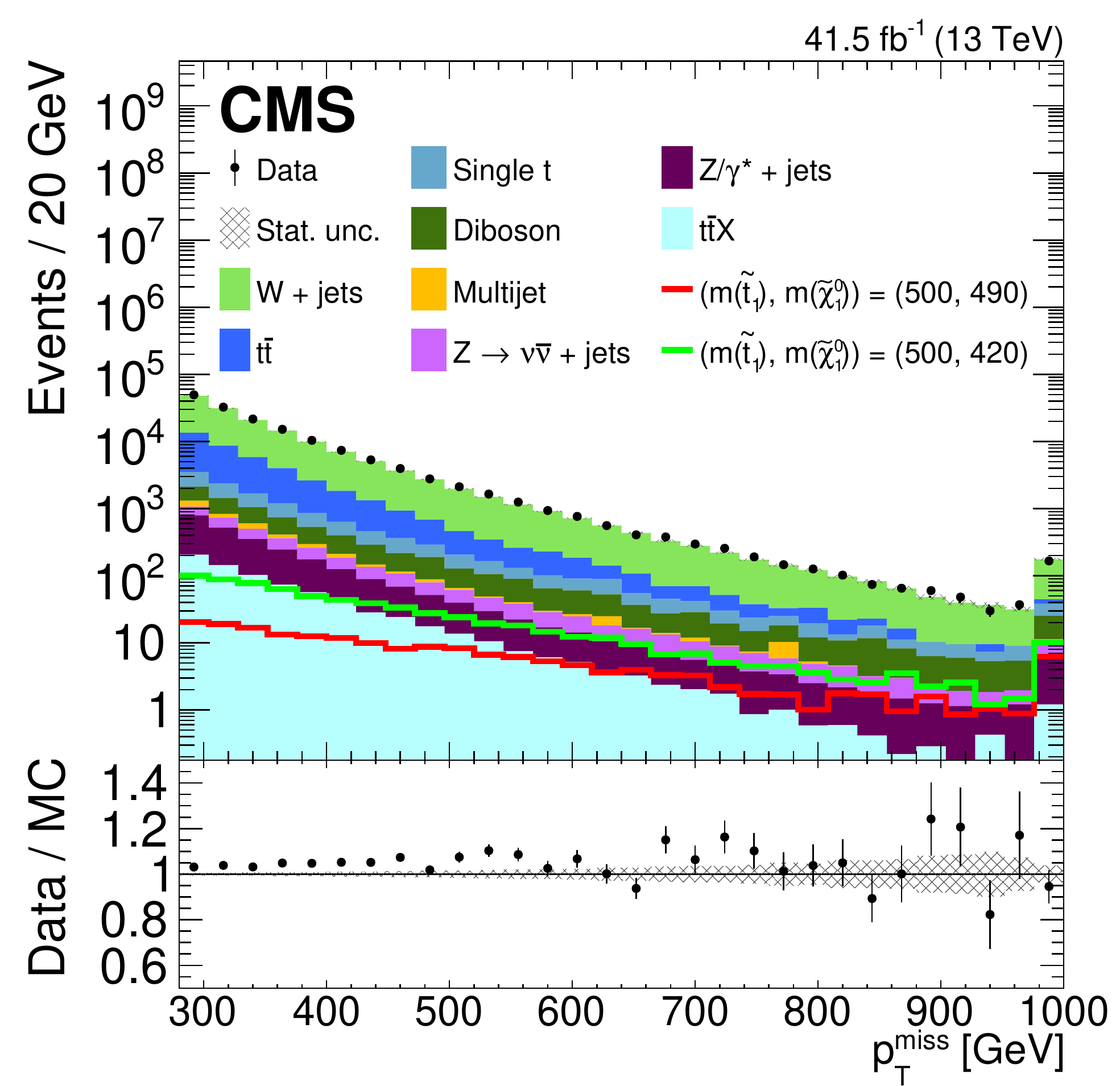} \hfil
\includegraphics[width=0.45\textwidth]{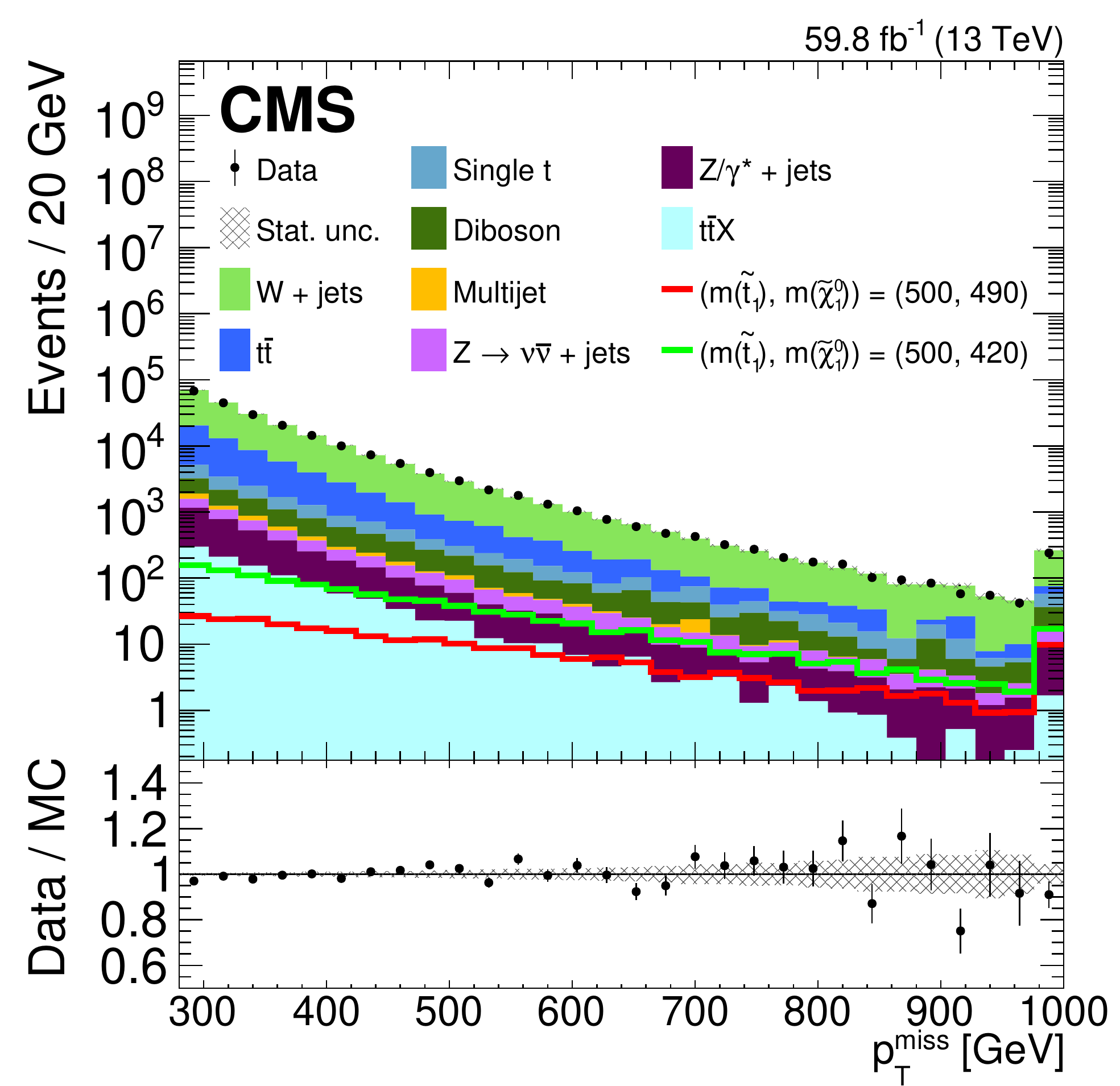} \\
\includegraphics[width=0.45\textwidth]{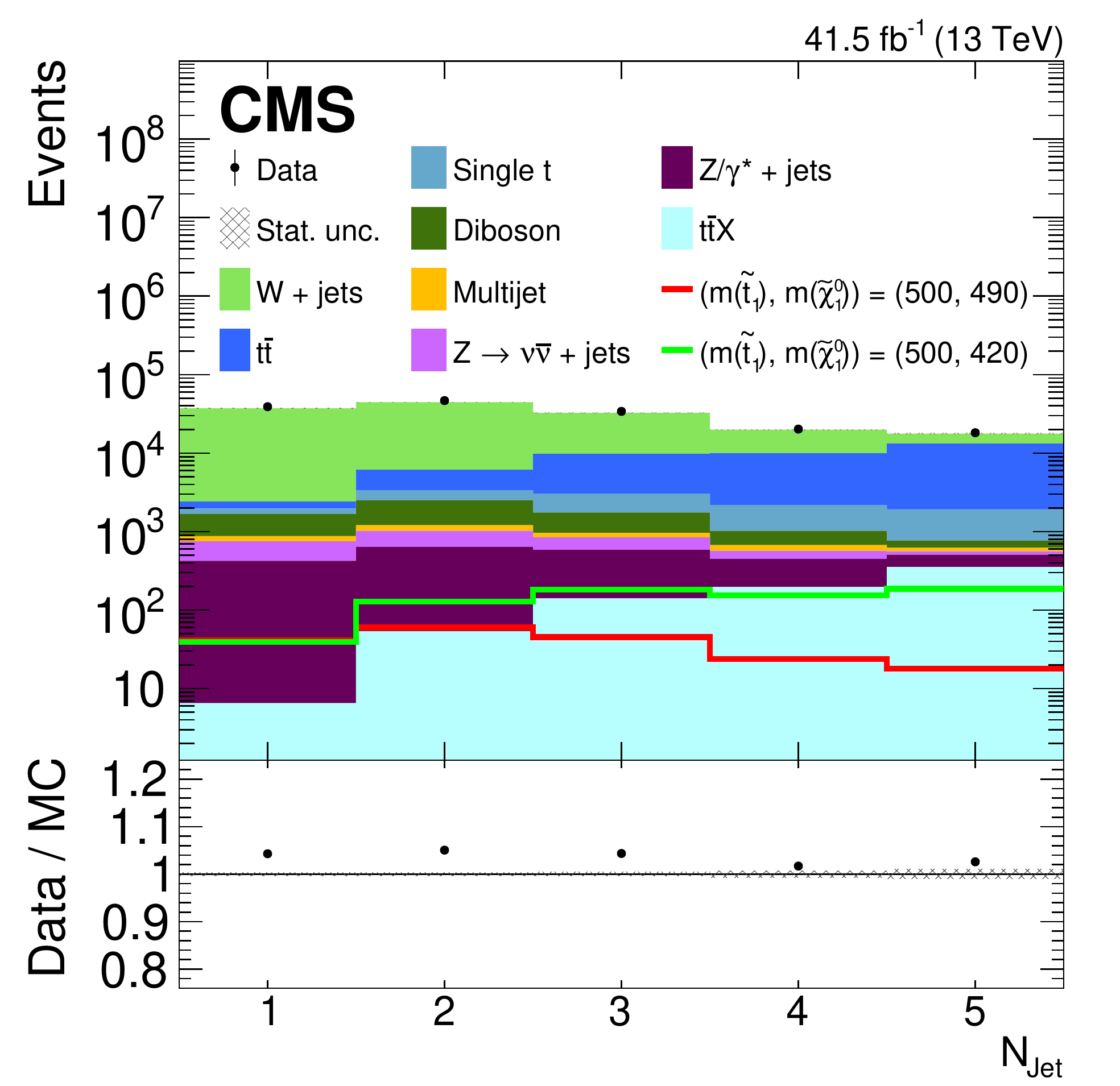} \hfil
\includegraphics[width=0.45\textwidth]{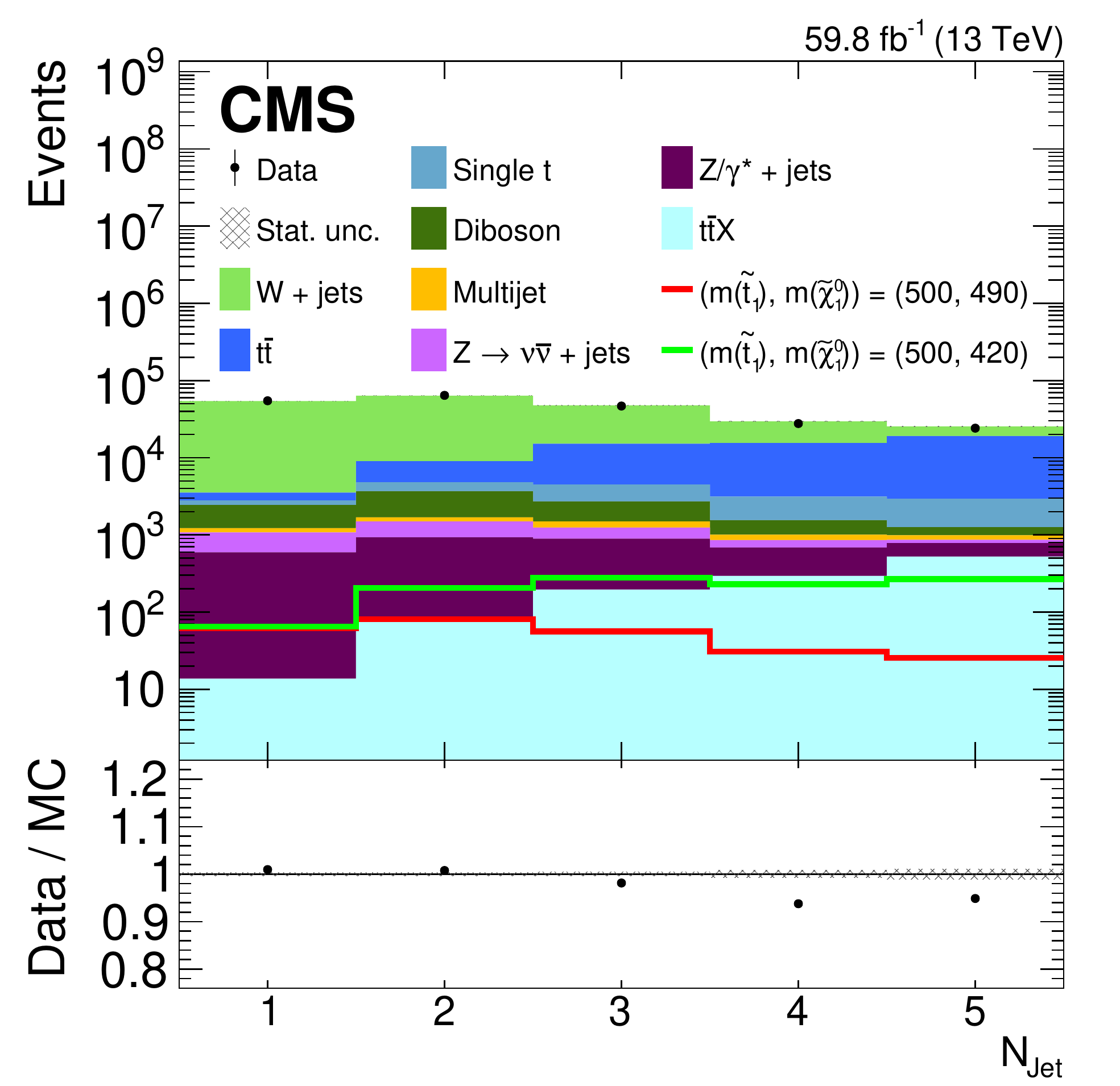}
\caption{Distributions of \ptl (upper), \ptmiss (middle), and \njet (lower), 
    after the preselection from 2017 (left) and 2018 (right) data (points) 
    and simulation (colored histograms).
    The simulated distribution of two signal points are represented by colored 
    lines, while not being stacked on the background distributions: 
    \mmplane = (500, 490) and (500, 420)\GeV. The last	bin in each plot includes 
    the overflow events. The lower panels show the ratio of data to the sum of 
    the simulated SM backgrounds. The shaded bands indicate only the statistical 
    uncertainty in the simulation predictions.
  }
\label{fig:presel}
\end{figure*}

After the preselection, the \wjets and \ttbar processes are the main SM backgrounds,
making up about 70 and 20\%, respectively, of the total expected background.
The \zinv process contributes to the SM background by having jets, genuine 
\ptmiss, and a jet misidentified as a lepton. The remaining background processes are 
diboson, single top quark, Drell--Yan (DY), multijet, and \ttX production where 
\PX is a vector boson. These processes are a less-important part of the expected 
background because of having a smaller cross section, a lower acceptance, or both.
The \ptl, \ptmiss, and \njet distributions after the preselection from the 2017
and 2018 data and the simulations are shown in Fig.~\ref{fig:presel}, where \njet 
is the number of jets in the event satisfying the jet criteria. The simulated 
background distributions for each year are normalized to the corresponding integrated 
luminosity.
The level of agreement with data gives us confidence in training the BDTs with the 
simulated distributions for the second step in the event selection.

\subsection{Classification and final selection}
\label{s:sel}

The selection of the signal events is based on a BDT~\cite{Hocker:2007ht} to take 
advantage of the different correlations among the discriminating variables for the 
signal and background processes. For each event passing the preselection, the BDT 
discriminator value, henceforth referred to as the BDT output, is evaluated. 
If the discriminator value exceeds the determined threshold, the event is retained. 
The choice of the discriminating variables used as input to the BDT is made by 
maximizing a figure of merit (FOM)~\cite{cowan} that takes into account the 
statistical and systematic uncertainties in a selection. Various BDTs are trained
with different sets of discriminating variables, and a variable is included in 
the final set only if it significantly increases the FOM obtained for any 
selection using the BDT output.
The list of the twelve retained input variables and a short description 
of their signal and background distributions is as follows:
\begin{itemize}
\item Variables related to \ptmiss: \ptmiss and \mt, where \mt is the 
  transverse mass of the lepton + \ptvecmiss system, defined as: 
  $\mT = \sqrt{\smash[b]{2\pt^\ell\ptmiss[1-\cos(\Delta\phi_{\ell,\ptvecmiss})]}}$,
  where $\Delta\phi$ is the azimuthal angular 
  difference between the lepton \ptvec and \ptvecmiss.
  The \ptmiss distribution extends to higher values for the signal than for the backgrounds due to the
 two undetected LSPs in the signal decays. The \mt spectrum peaks around 80\GeV for the
 SM background and is a broad distribution for the signal.
\item Lepton-related variables: \ptl, \etl, and \chgl. 
The correlations between \ptmiss and \ptl are different for the signal, where \ptmiss comes
from three undetected particles (two \PSGczDo and a \PGn), than for \wjets and \ttbar 
backgrounds, where \ptmiss is the result of a single undetected particle (\PGn). 
Because the decay products of the signal are more centrally produced than those of the \wjets process, 
the lepton pseudorapidity \etl distribution is populated at more central values for
the signal than this background.
  The lepton charge \chgl is a discriminating variable because \PWp 
  and \PWm bosons are not produced equally at the LHC, while the signal events contain 
  equal numbers of positively and negatively charged leptons.
\item Jet-related variables: \ptisr, \ptb, \njet, and \HT. The variable
  \ptisr is defined as the \pt of the leading jet, and selects the high-momentum ISR jet in 
  signal events.
  The \ptb variable is the transverse momentum of the jet with the highest \PQb tagging discriminant value.
  Both the \ptisr and \ptb 
  variables are sensitive to the available phase space, which depends on $m(\stp)-m(\lsp)$ for the signal, and 
  $m(\cPqt)-m(\PW)$ for the \ttbar background. The \njet variable is sensitive to the mass difference \dm,
  while the \HT variable provides discrimination between
  signal and both the \wjets and \ttbar backgrounds.
\item \PQb jet-related variables: \nbl, \drLB, and \bdisc.
  The number of loosely \PQb tagged jets 
  \nbl, the distance in ($\eta$, $\phi$) space between the directions of the lepton and the 
  jet with the highest \PQb tagging 
  discriminant \drLB, and the highest \PQb tagging discriminant per event \bdisc are
  included as input variables. 
  While the preselection has no requirement on 
  \PQb tagging, information related to it is passed to the BDT to help discriminate between the signal and mainly
  the \wjets background.
\end{itemize}
The five most discriminating variables, in decreasing power, are \ptl, 
\ptmiss, \ptisr, \HT, and \mt.

\begin{figure}[!htbp]
\centering
\includegraphics[width=0.45\textwidth]{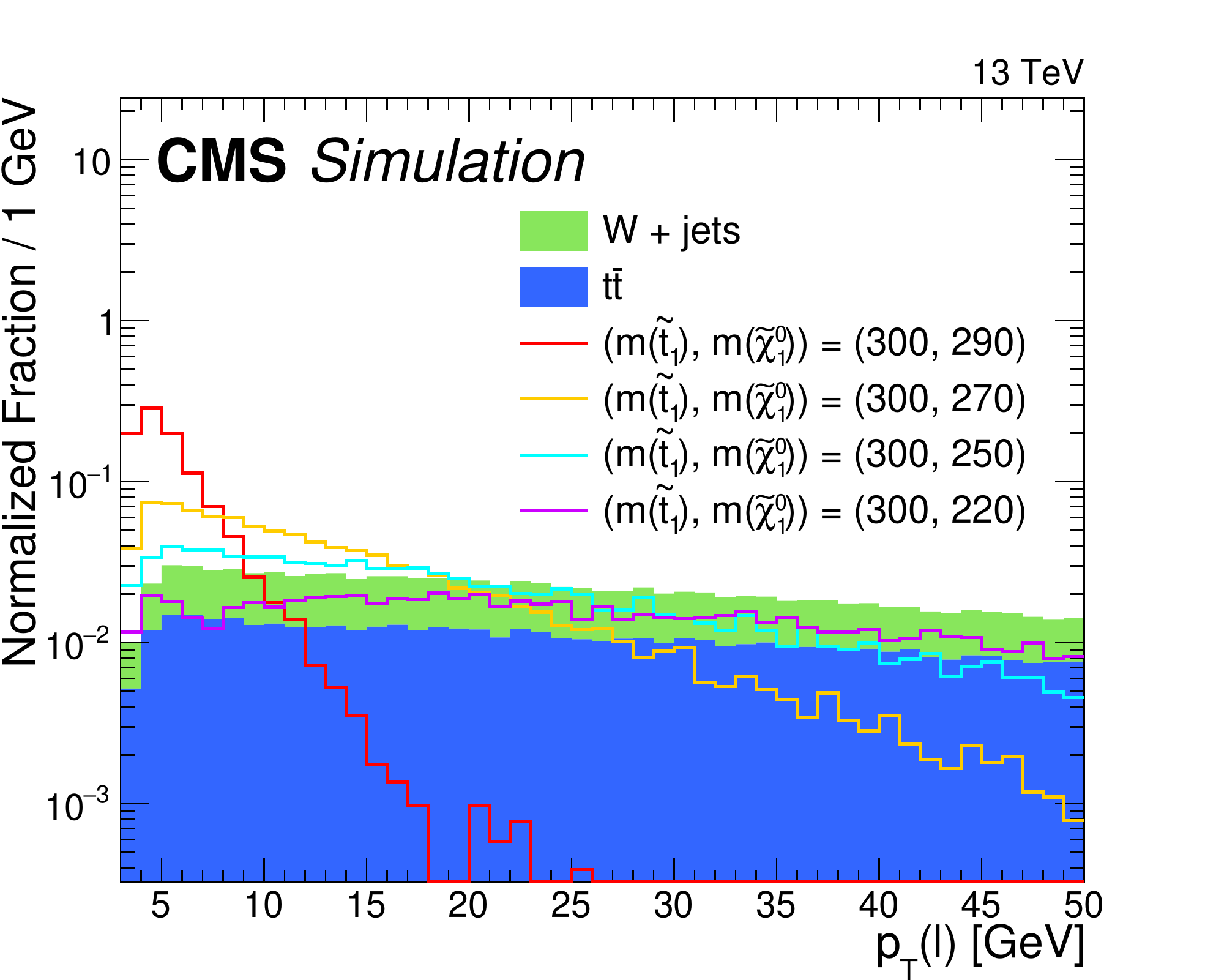} \hfil
\includegraphics[width=0.45\textwidth]{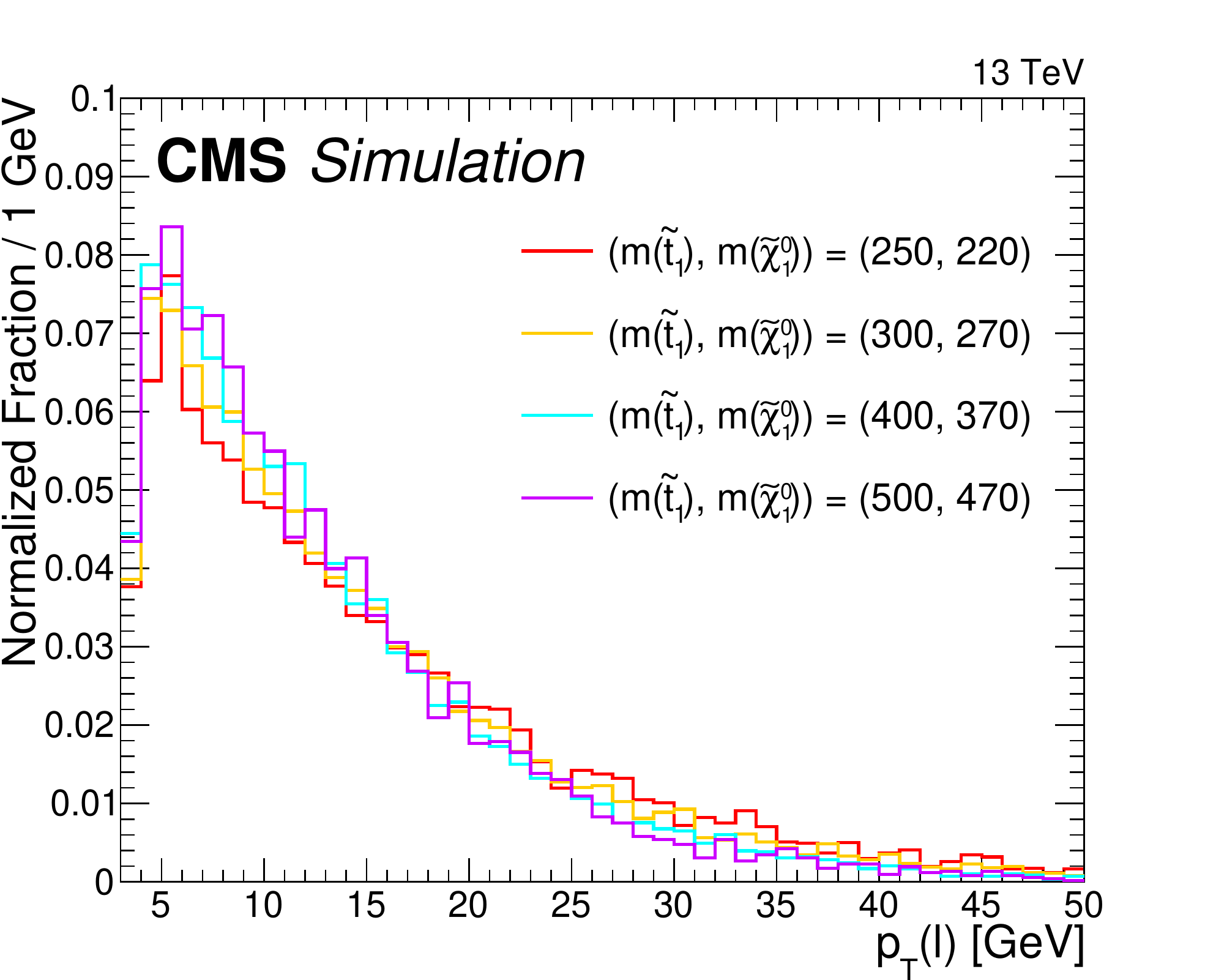} \\
\includegraphics[width=0.45\textwidth]{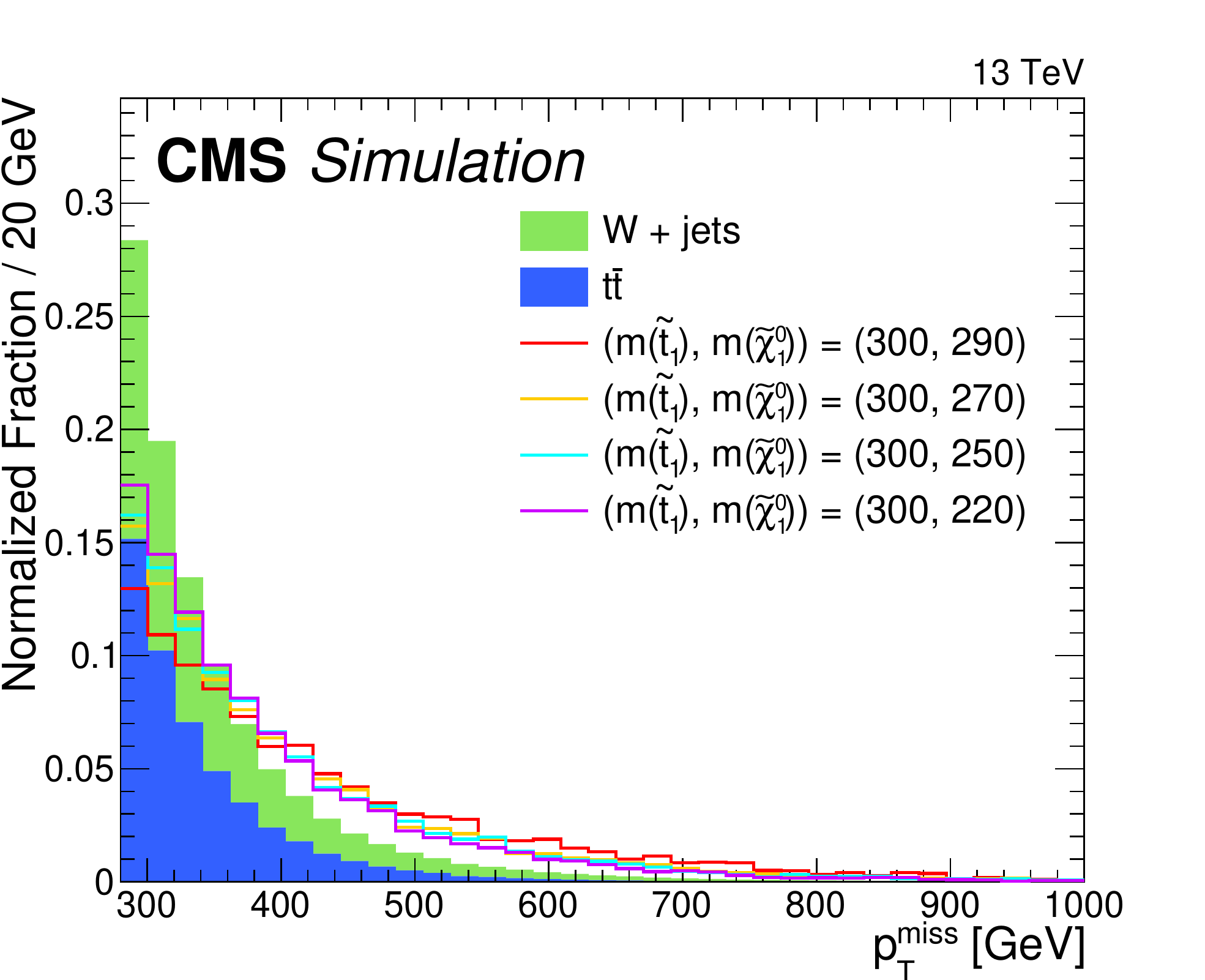} \hfil
\includegraphics[width=0.45\textwidth]{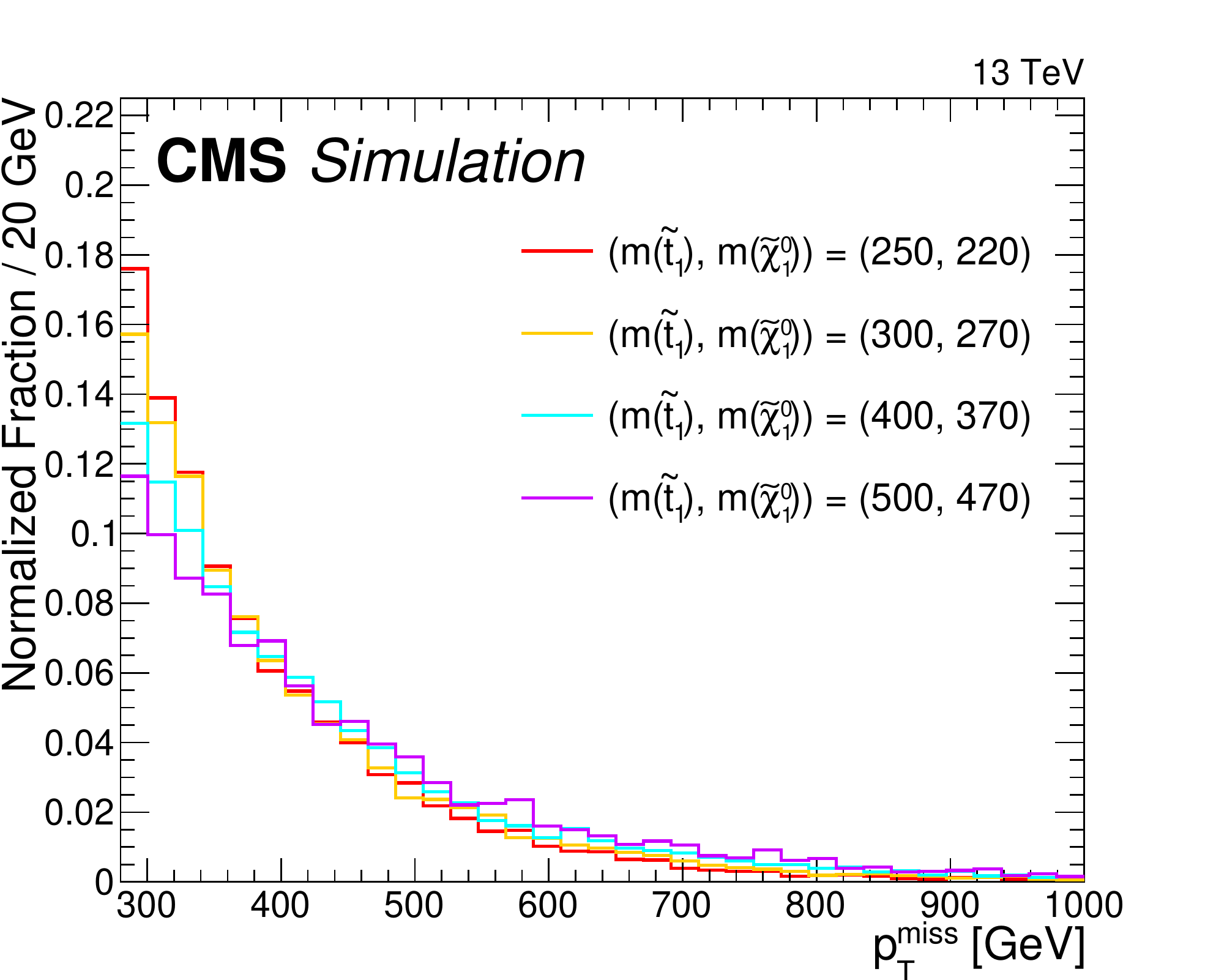} \\
\includegraphics[width=0.45\textwidth]{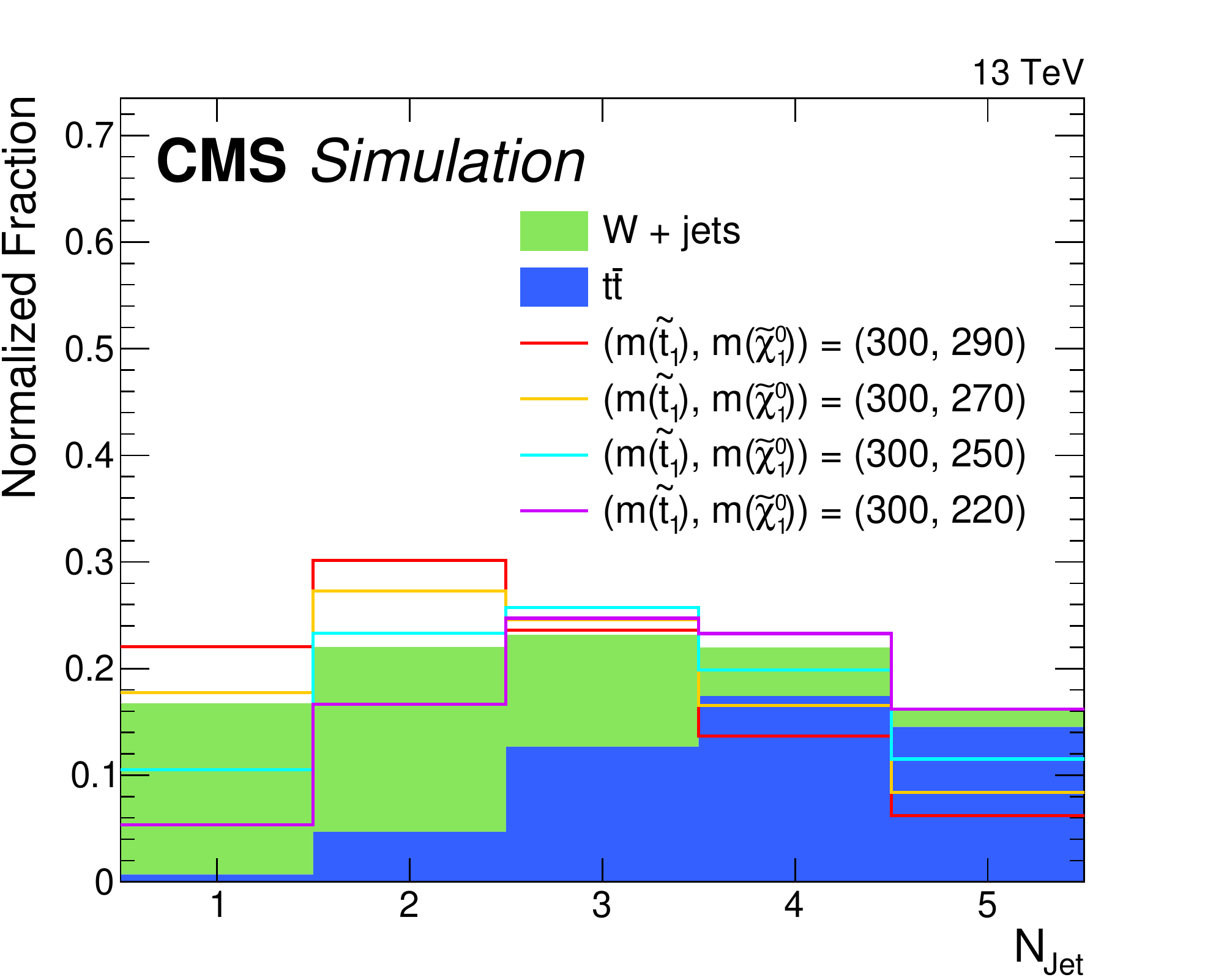} \hfil
\includegraphics[width=0.45\textwidth]{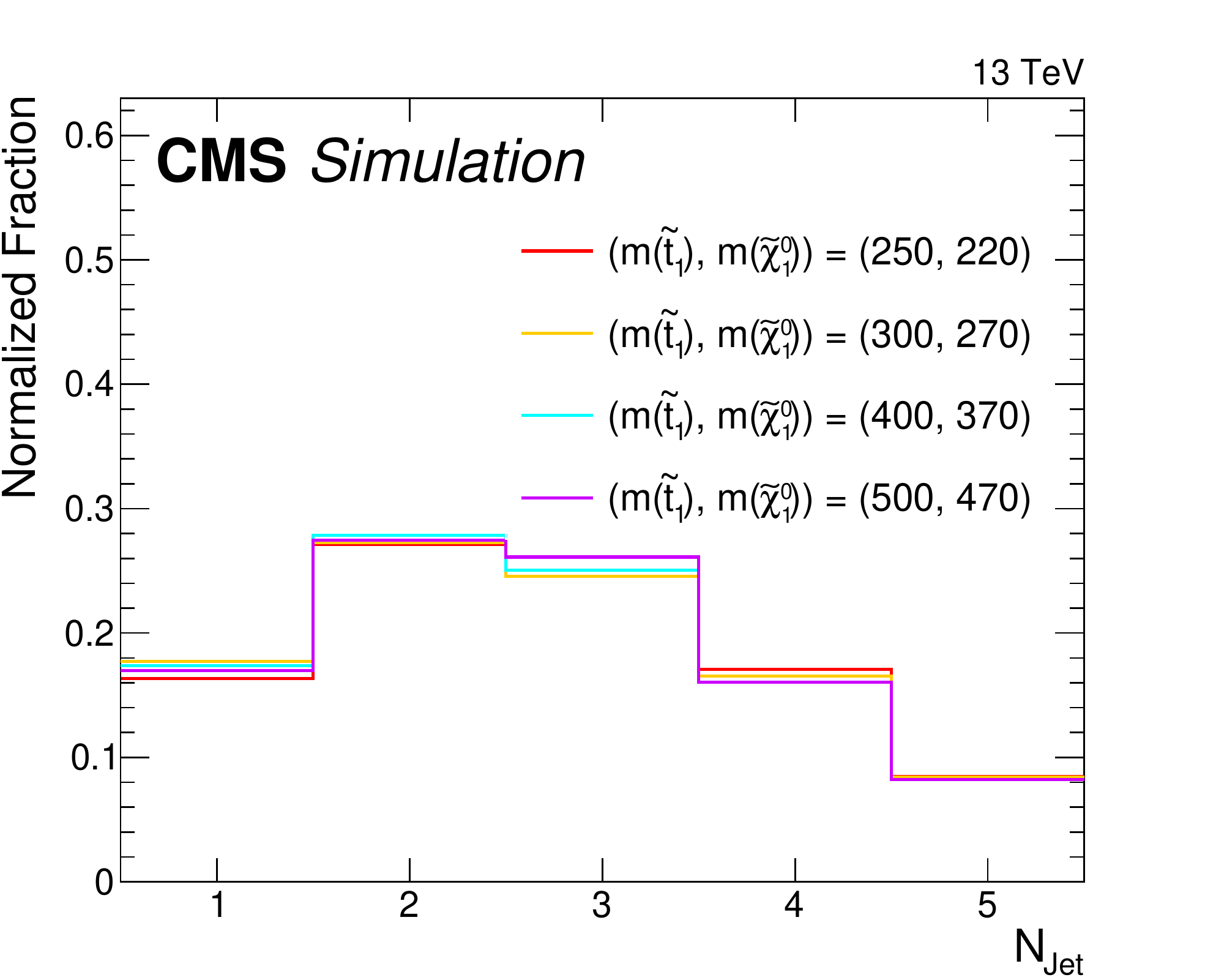} 
\caption{Simulated distributions of \ptl (upper), \ptmiss (middle), and 
\njet (lower) after the preselection.
The \wjets and \ttbar background distributions are shown as colored histograms, 
and the signal distributions by the solid lines.
The total background distribution and the signal distributions are all normalized to unit area.
On the left, the signal distributions are given for a top squark mass of 300\GeV and $\dm=10$, 30,
50, and 80\GeV.
On the right, the signal distributions are shown for four different \mmplane 
values, all corresponding to the same $\dm=30\GeV$.
}
\label{fig:varshape}
\end{figure}

The discrimination power of the input variables varies as a function of \dm,
as illustrated in Fig.~\ref{fig:varshape} (left). An important feature of 
this search is the adaptation of the selection tool to the evolving kinematic 
variables of the signal over the \mmplane plane. Therefore, this plane is divided
into eight \dm regions (from 10 to 80\GeV, in steps of 10), and a separate BDT is
trained for each \dm region. 
The BDTs are trained to discriminate signal from background using the binomial 
log-likelihood loss function. Only the \wjets and \ttbar processes, which 
constitute a large fraction of the total background after preselection, are 
included in the training. They are normalized in proportion to their theoretical
cross sections.
As seen in Fig.~\ref{fig:varshape}~(right), different signal 
points with the same \dm have similar input variable distributions. 
This is expected since with the same \dm they have the same available phase space.
Because of this, all the signal points with the same \dm are grouped together when
training the BDT, thus increasing the number of signal events for each training. Because
of the large variation of the \ptl spectrum across the \mmplane plane, $\ptl < 30\GeV$
is required for $\dm < 70\GeV$ signal regions before training the BDTs, while 
imposing no restriction on \ptl for signal regions with higher \dm. This improves
the ability of the BDT to separate the signal from the \ttbar background.

\begin{figure}[!htbp]
\centering
\includegraphics[width=0.32\textwidth]{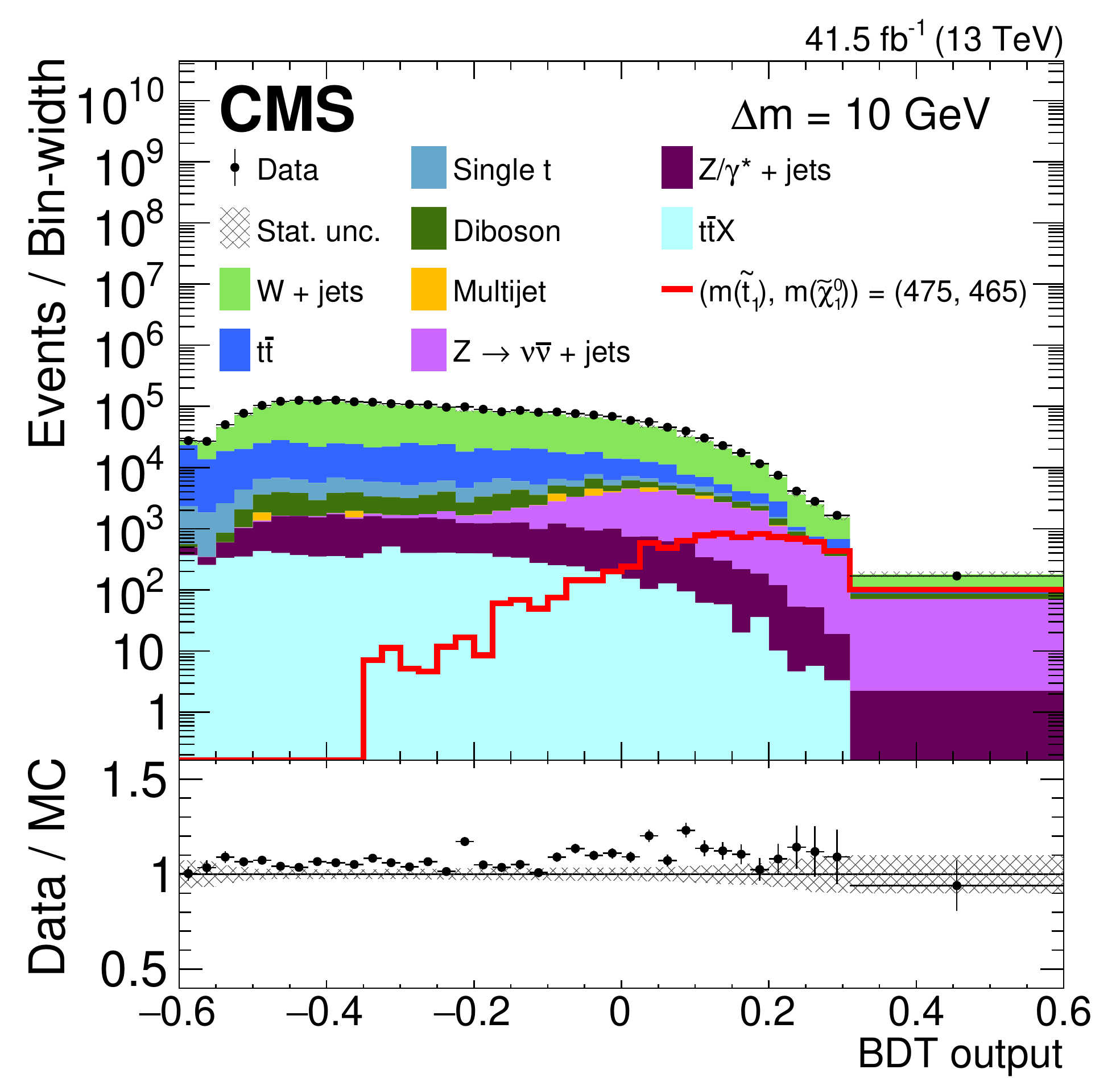} \hfil
\includegraphics[width=0.32\textwidth]{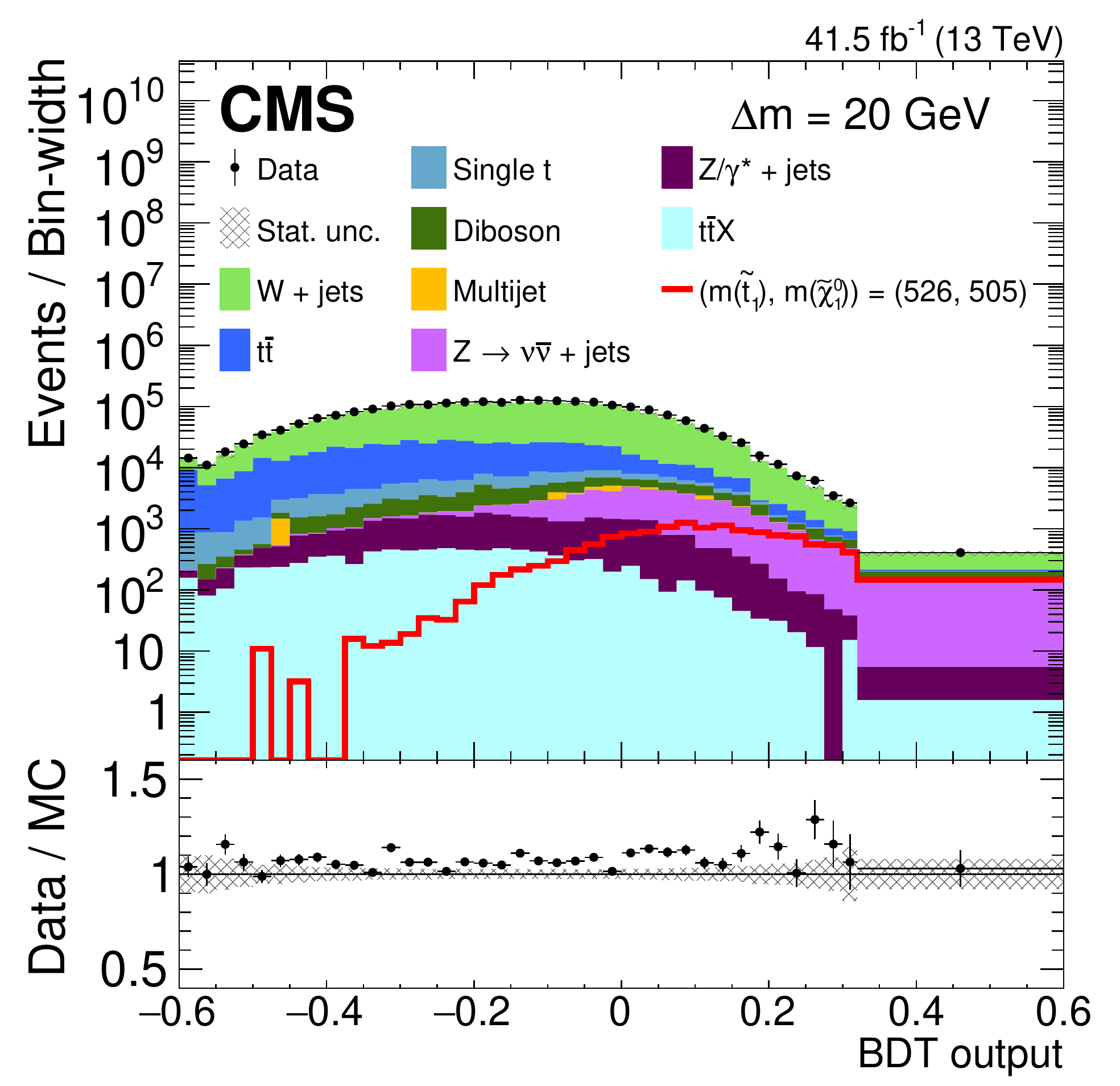} \hfil
\includegraphics[width=0.32\textwidth]{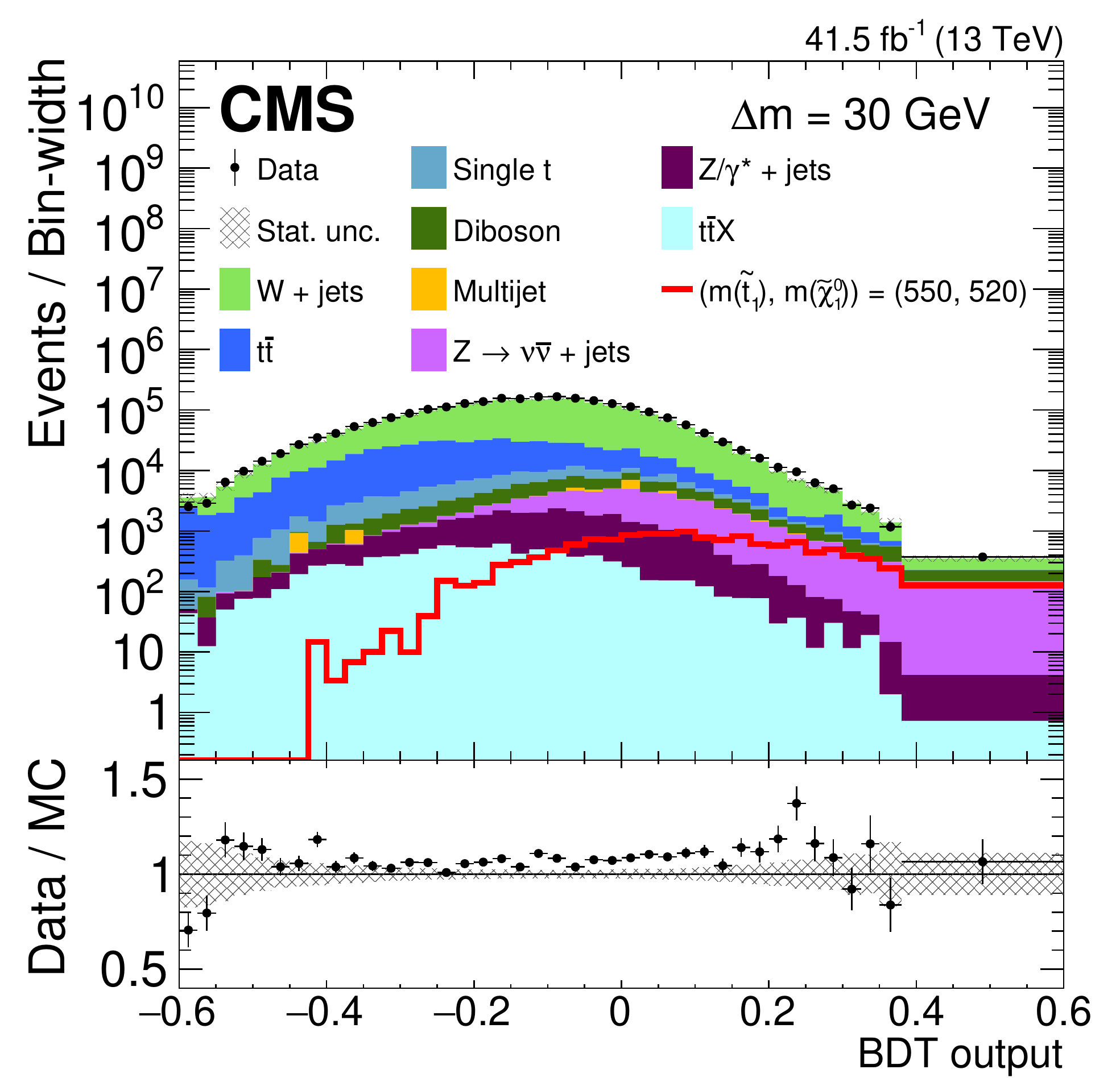} \\
\includegraphics[width=0.32\textwidth]{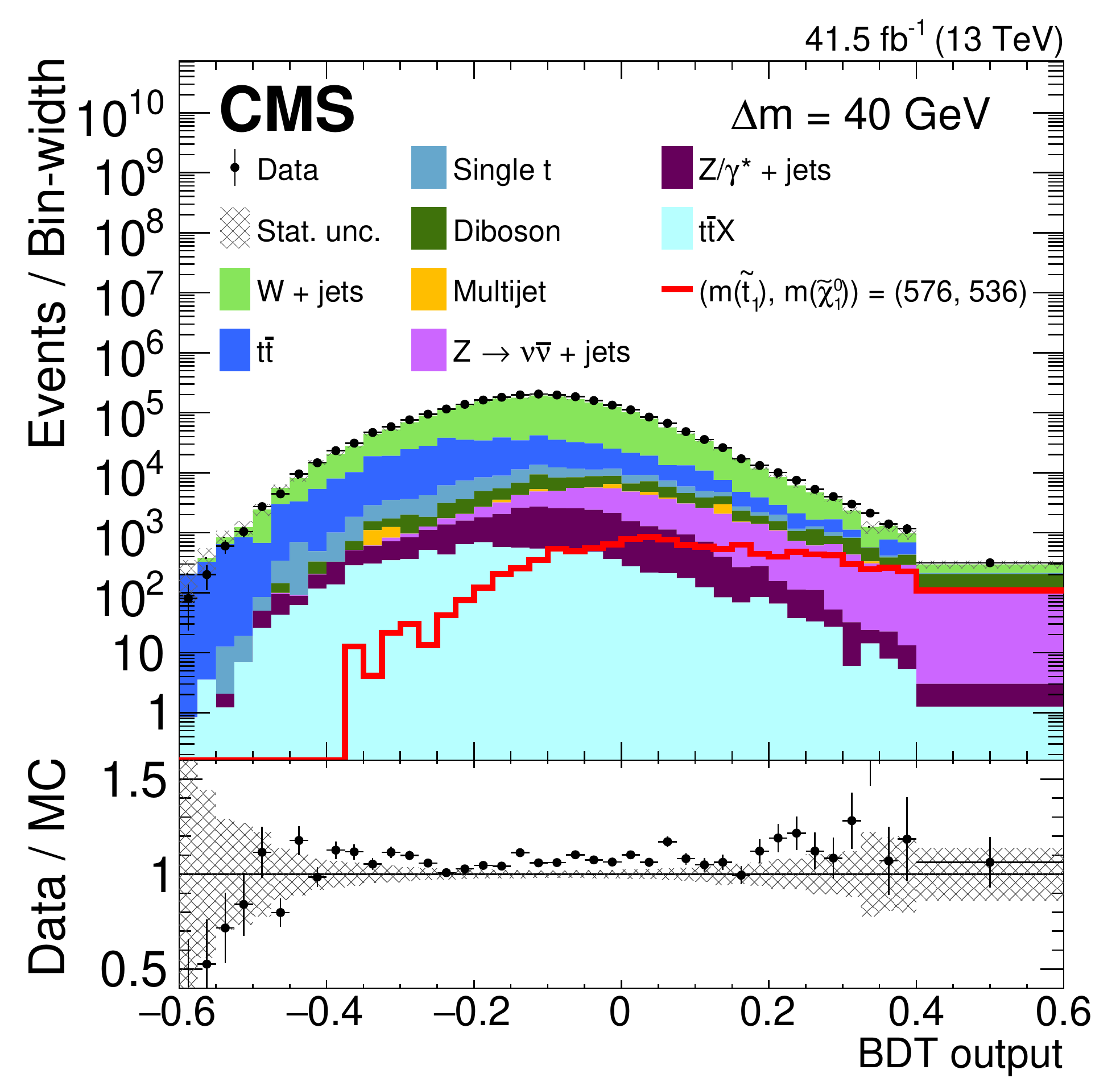} \hfil
\includegraphics[width=0.32\textwidth]{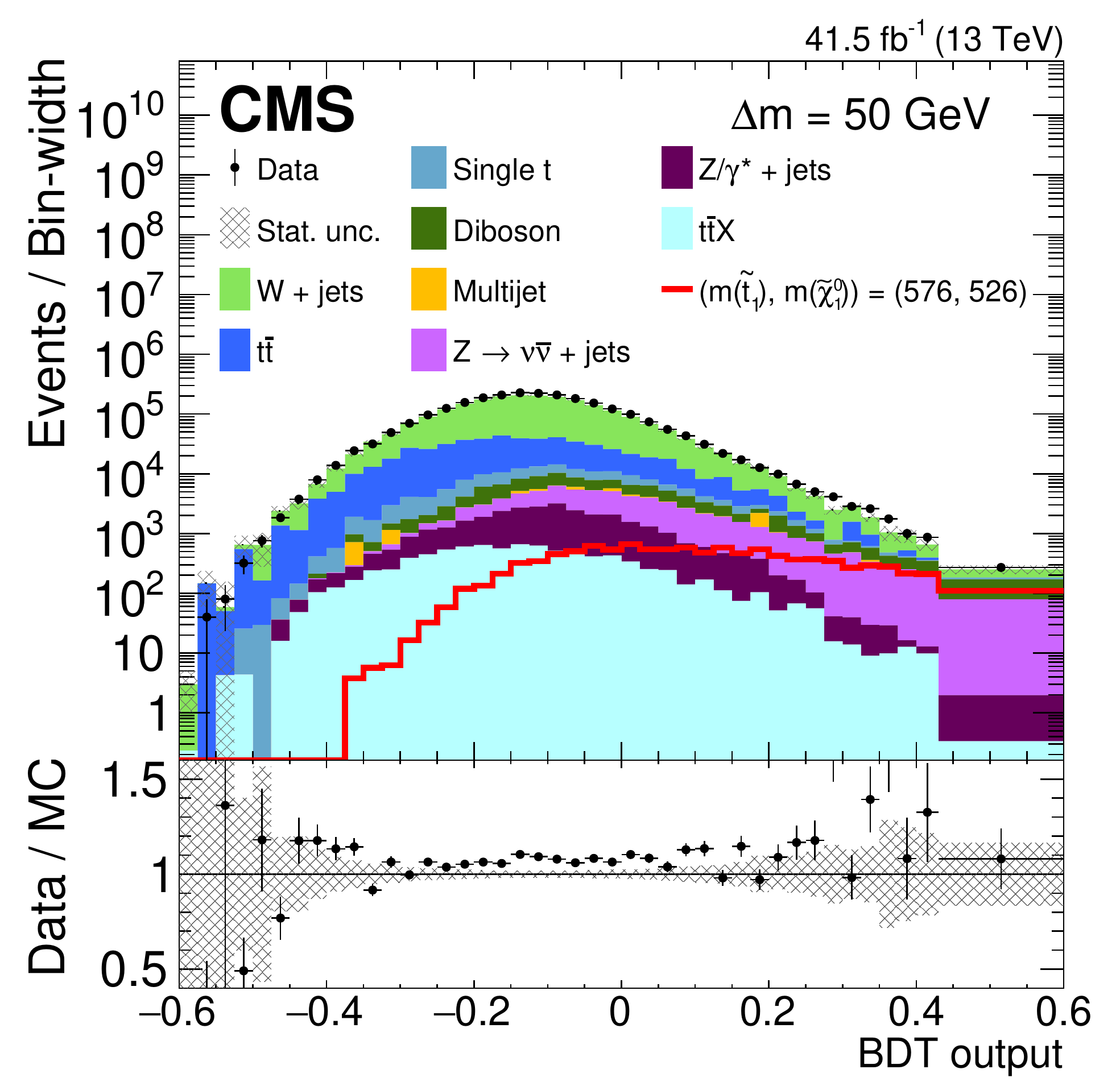} \hfil
\includegraphics[width=0.32\textwidth]{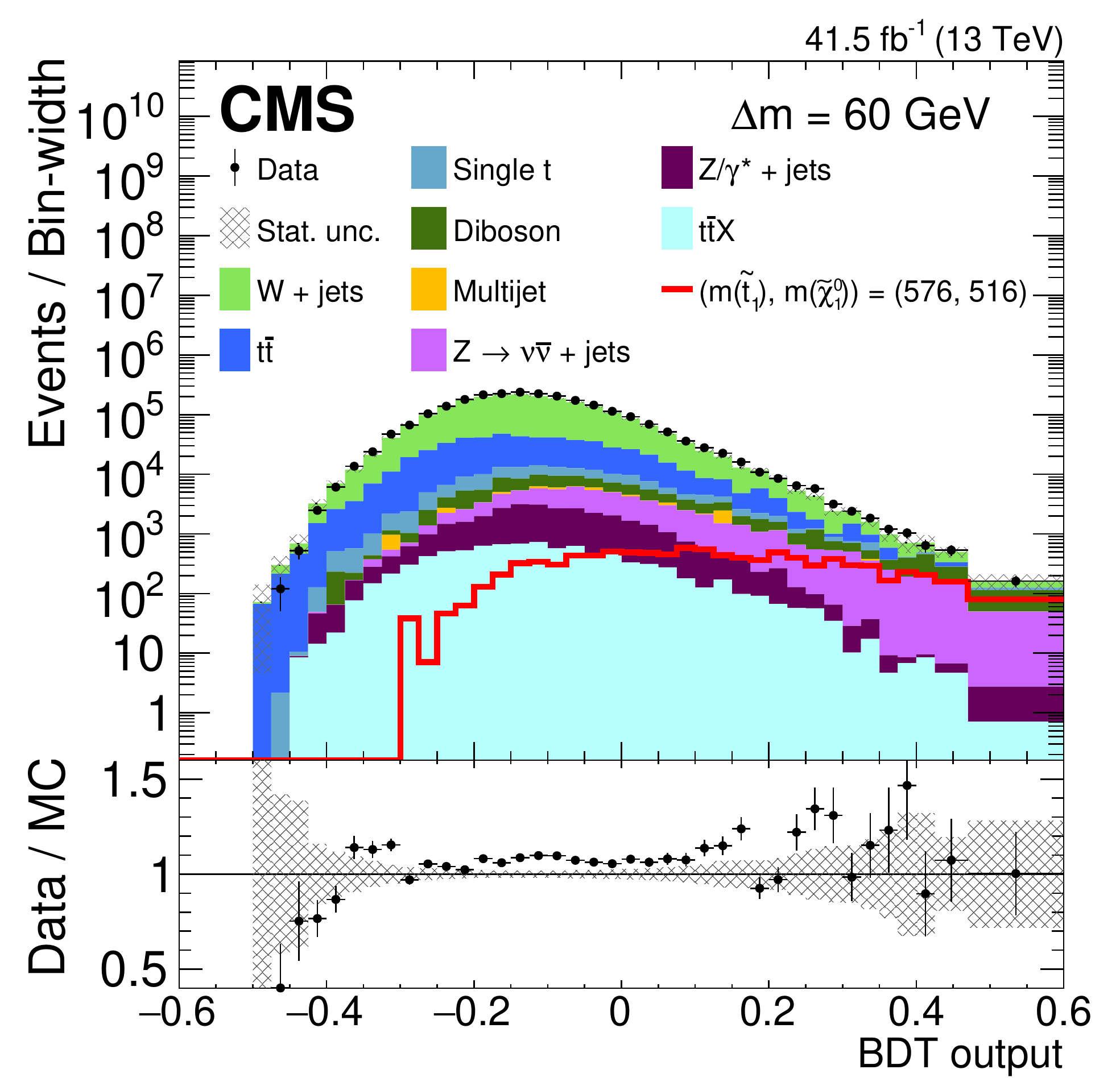} \\
\includegraphics[width=0.32\textwidth]{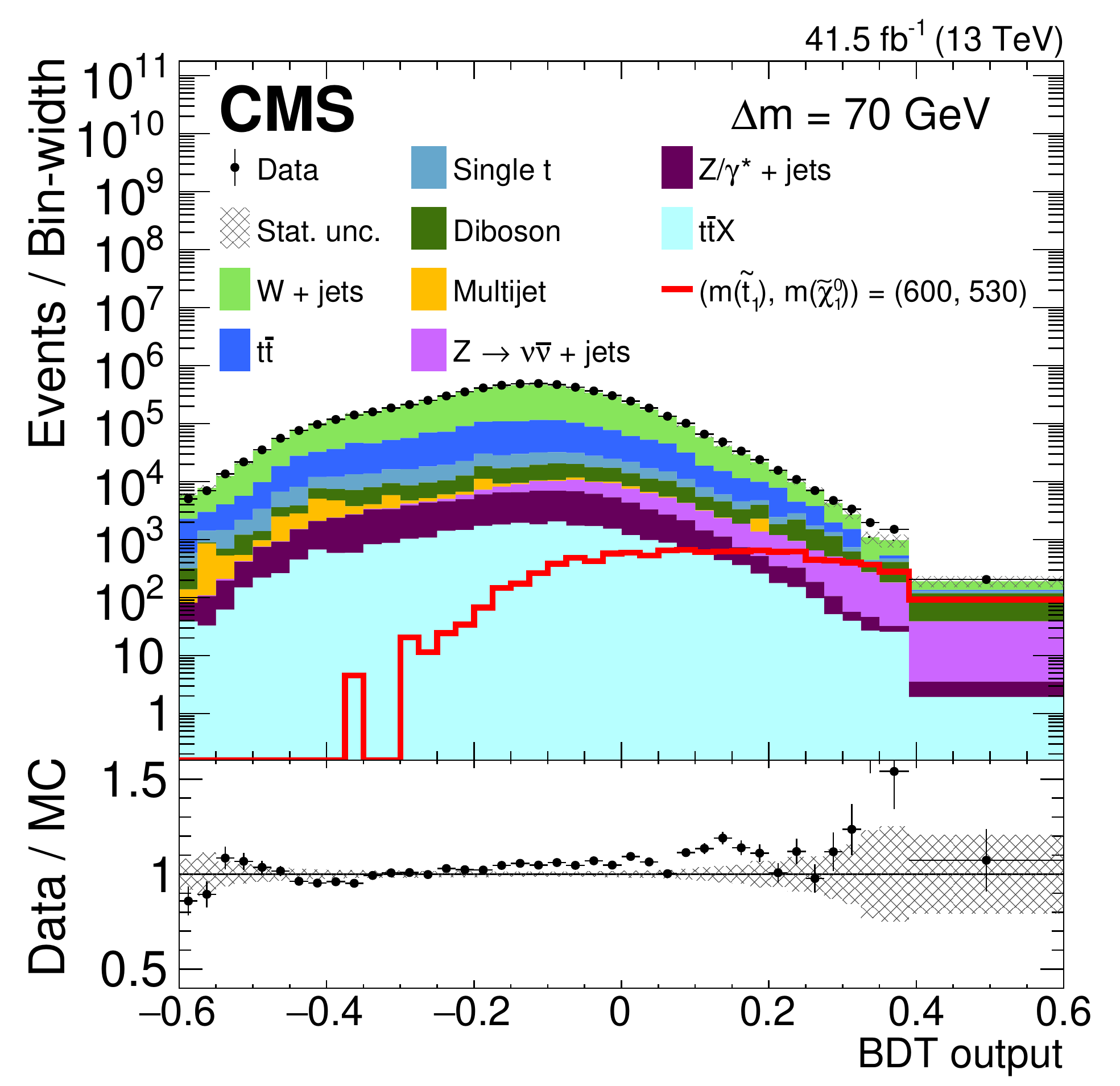} \hfil
\includegraphics[width=0.32\textwidth]{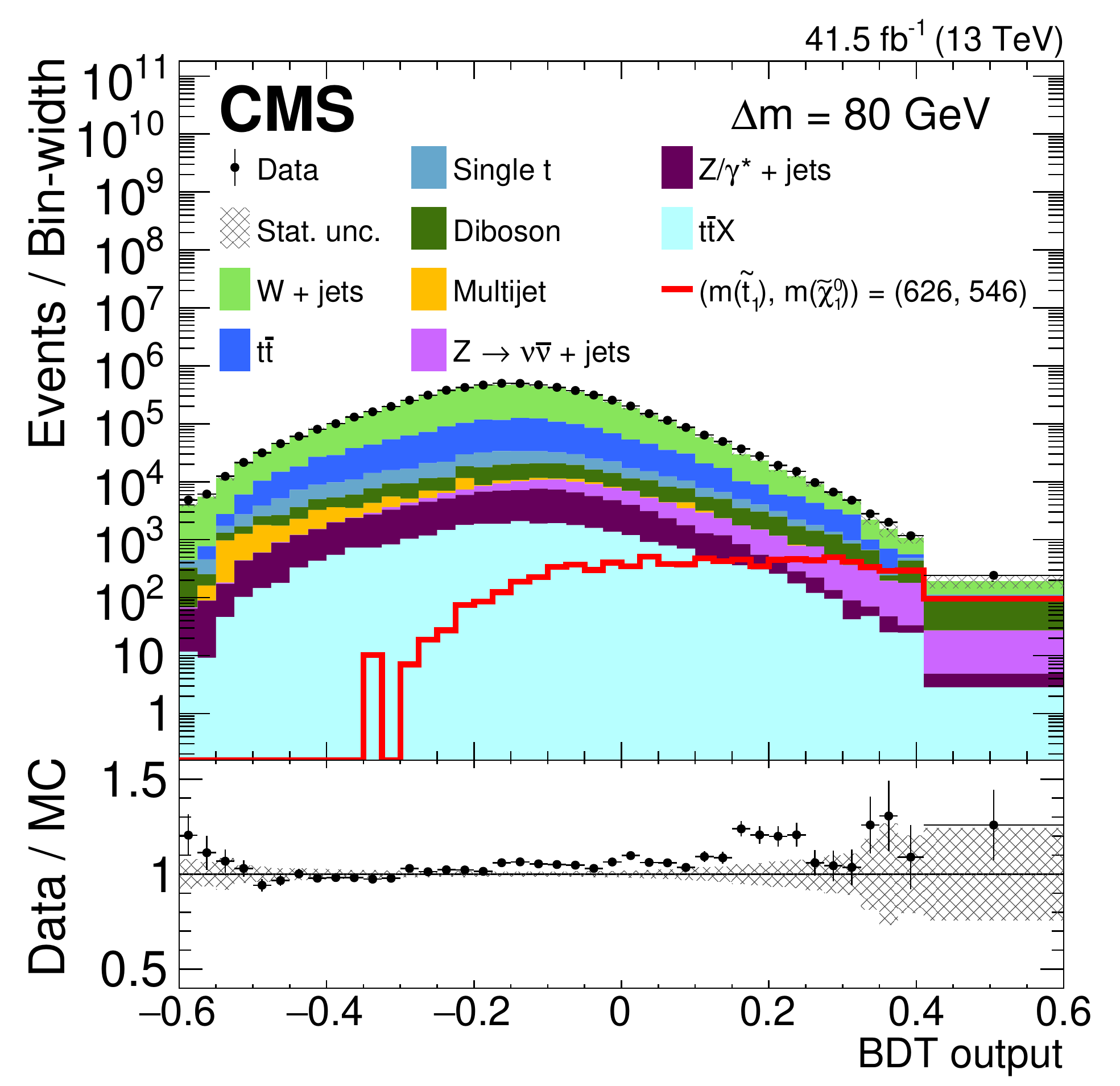} \hfil
\caption{BDT output distributions from data (points) and simulation (colored histograms) 
        after the preselection in 10\GeV steps of \dm from 10 (upper left) to 80 (lower 
        right)\GeV for the 2017 data. The last bin corresponds to the SR. 
        For each \dm value, the predicted signal distribution is shown by the
        solid red line for a representative \mmplane point, unstacked from the 
        histograms. The lower panels show the ratio of the data to the sum of the
        background predictions, with the vertical bars and shaded area giving 
        only the statistical uncertainty in the data and the simulated background, 
        respectively.
}
\label{fig:bdtdatamc1}
\end{figure}

\begin{figure}[!htbp]
\centering
\includegraphics[width=0.32\textwidth]{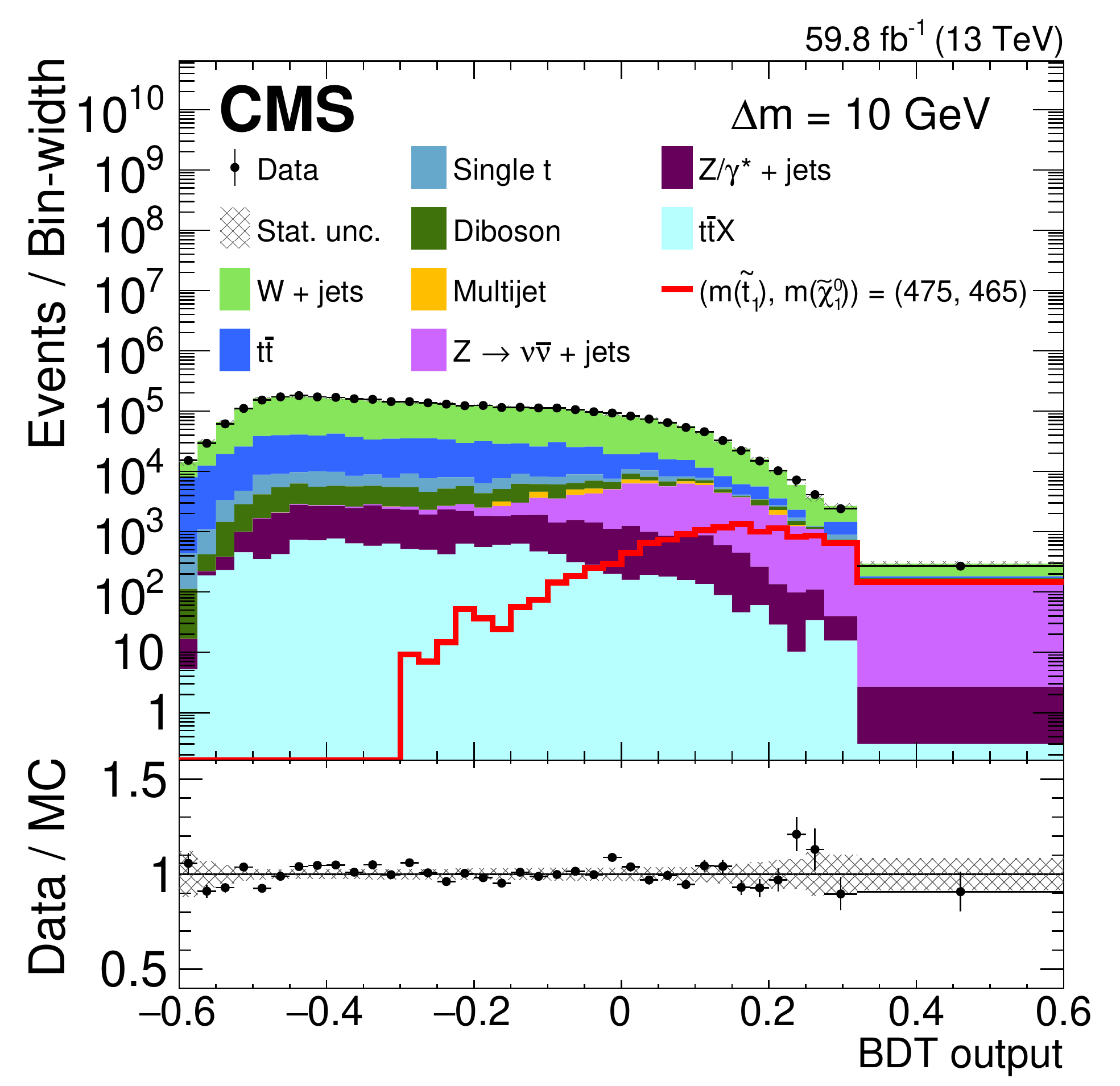} \hfil
\includegraphics[width=0.32\textwidth]{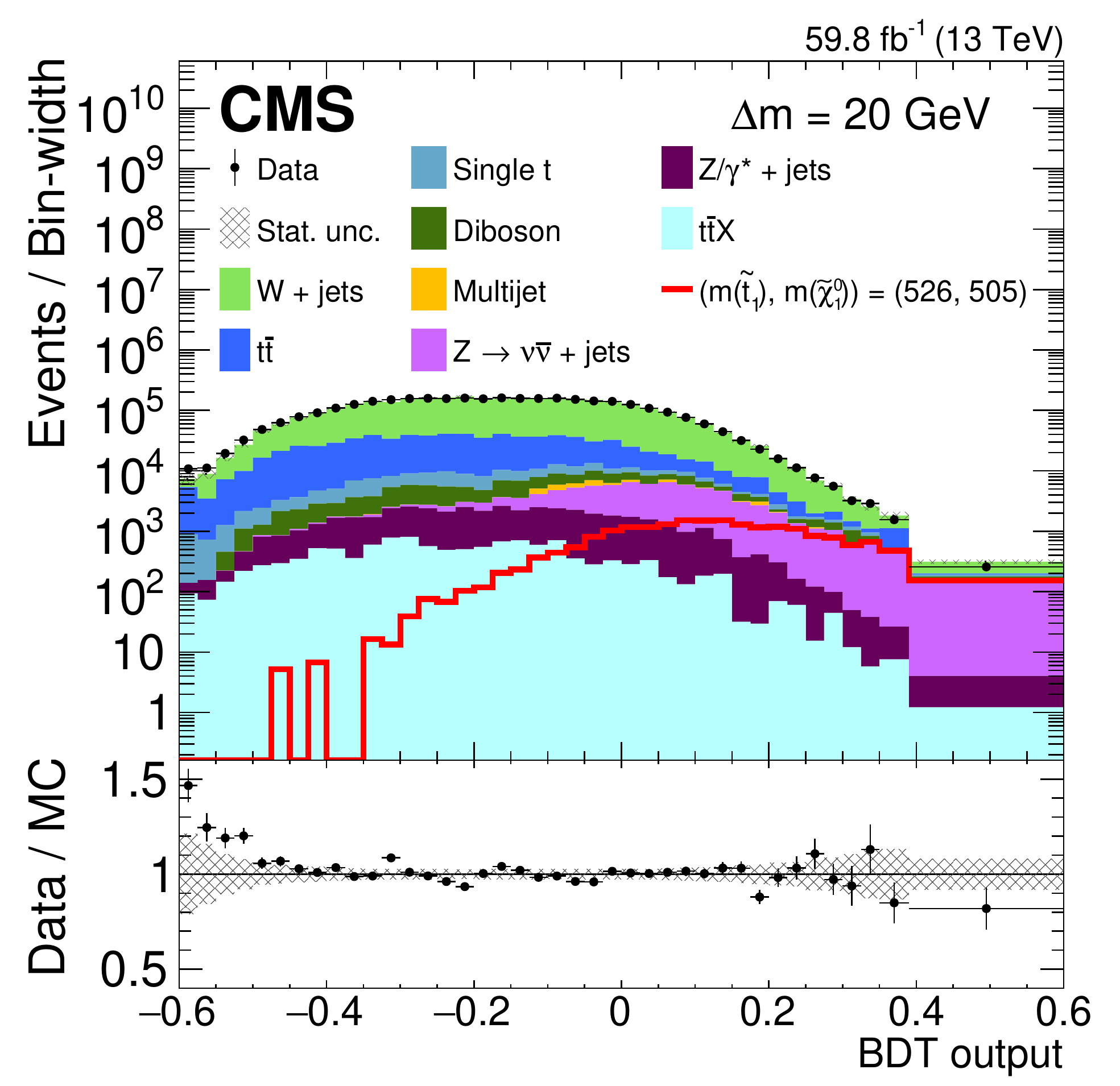} \hfil
\includegraphics[width=0.32\textwidth]{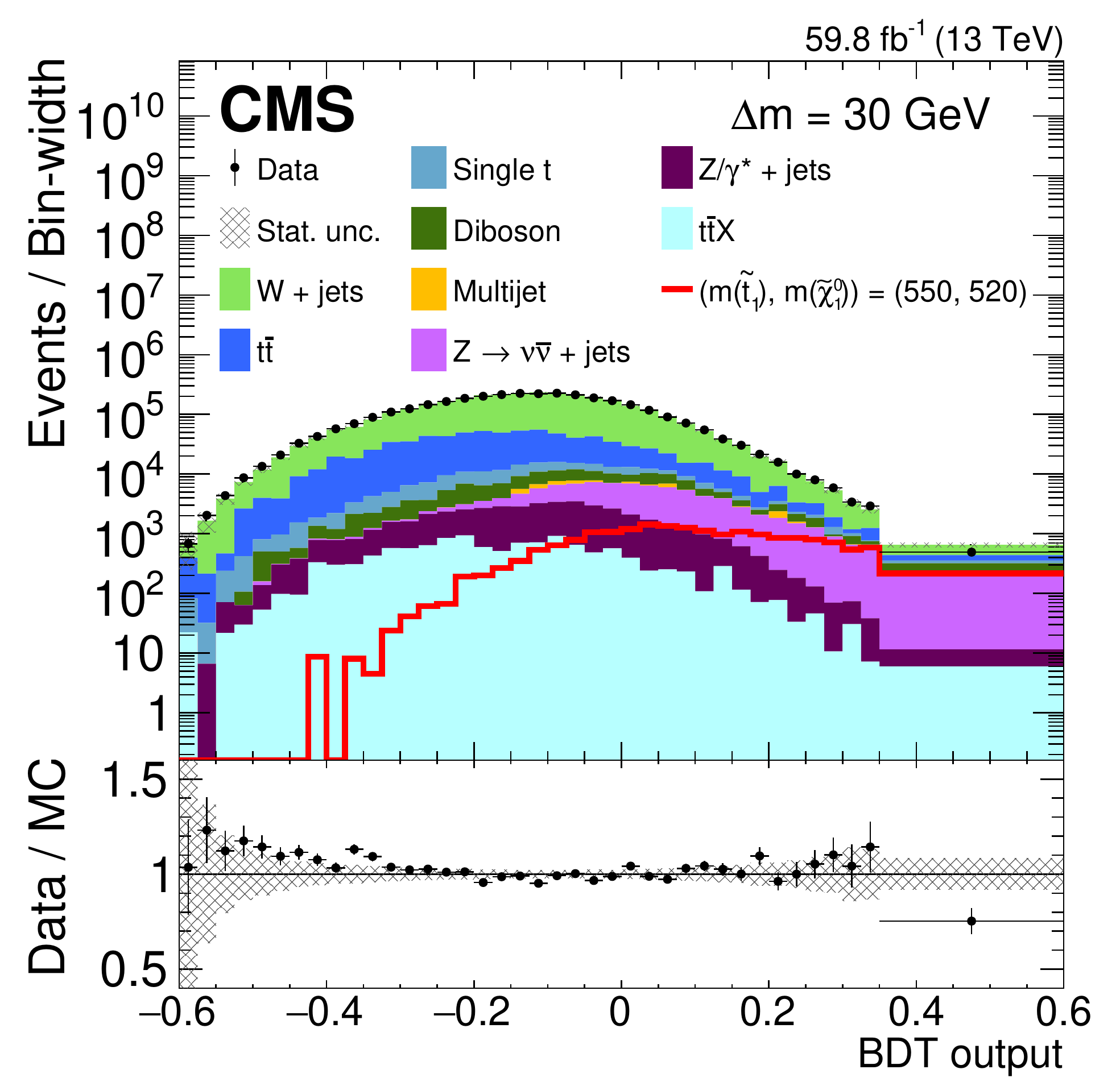} \\
\includegraphics[width=0.32\textwidth]{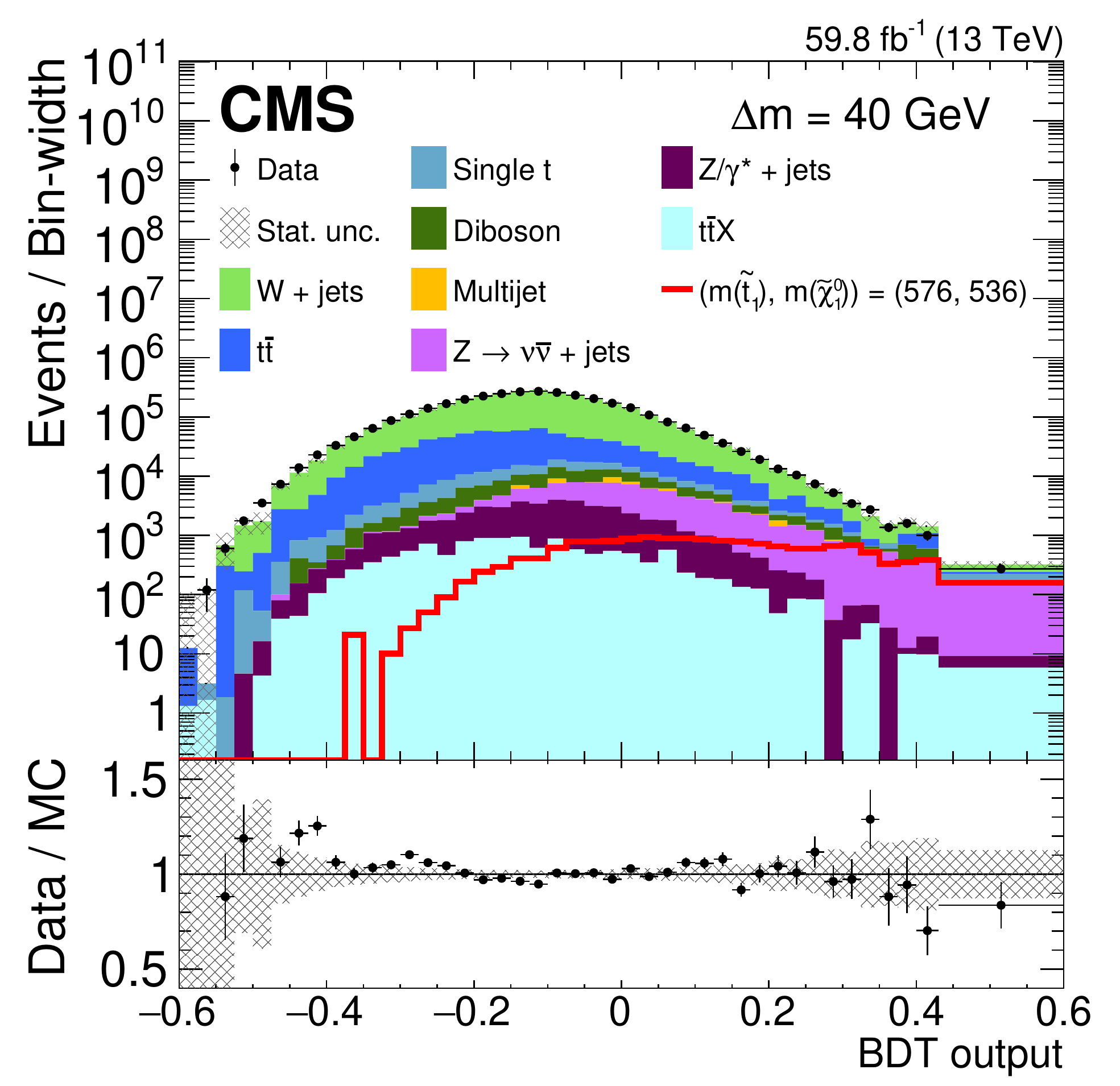} \hfil
\includegraphics[width=0.32\textwidth]{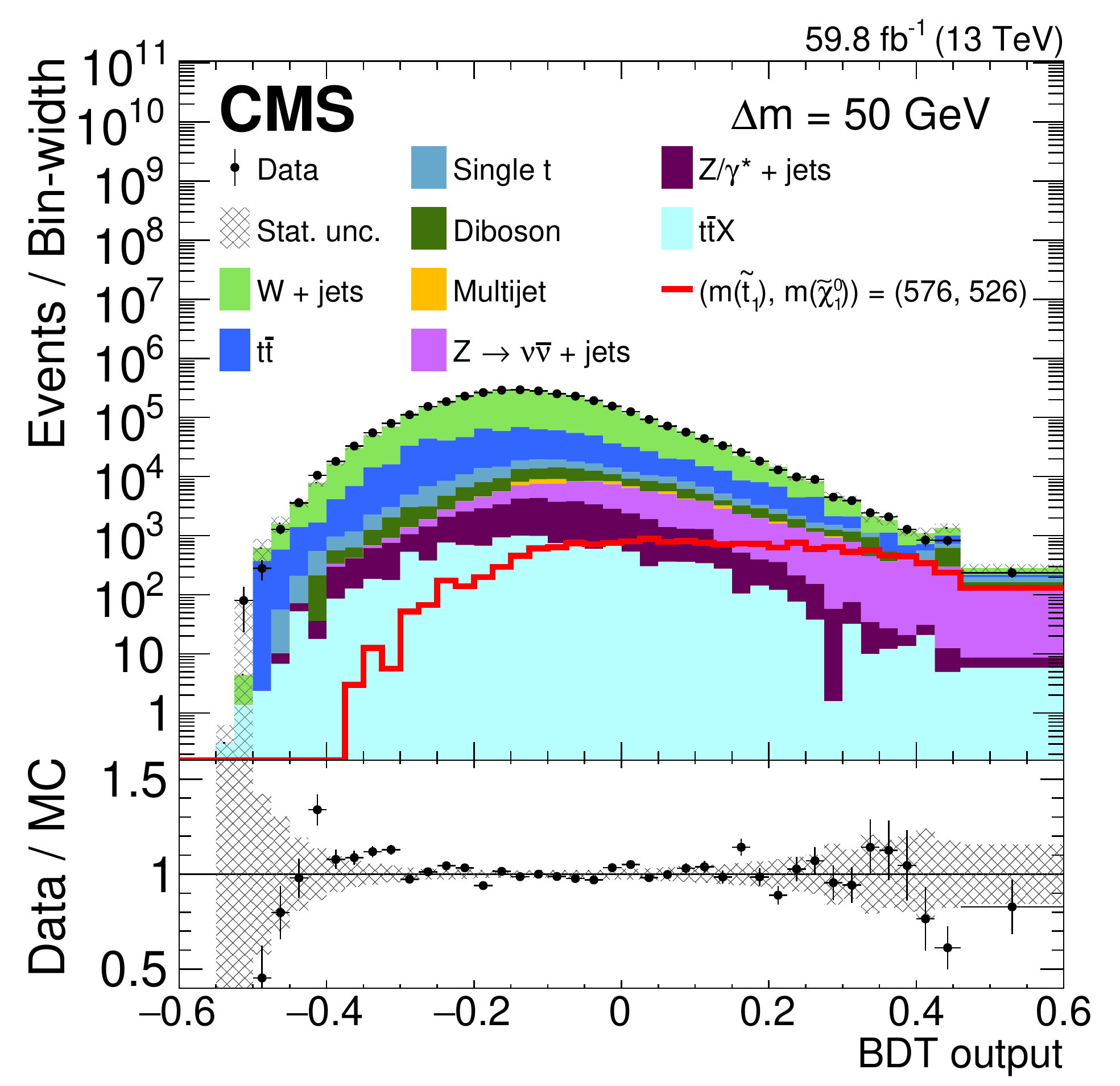} \hfil
\includegraphics[width=0.32\textwidth]{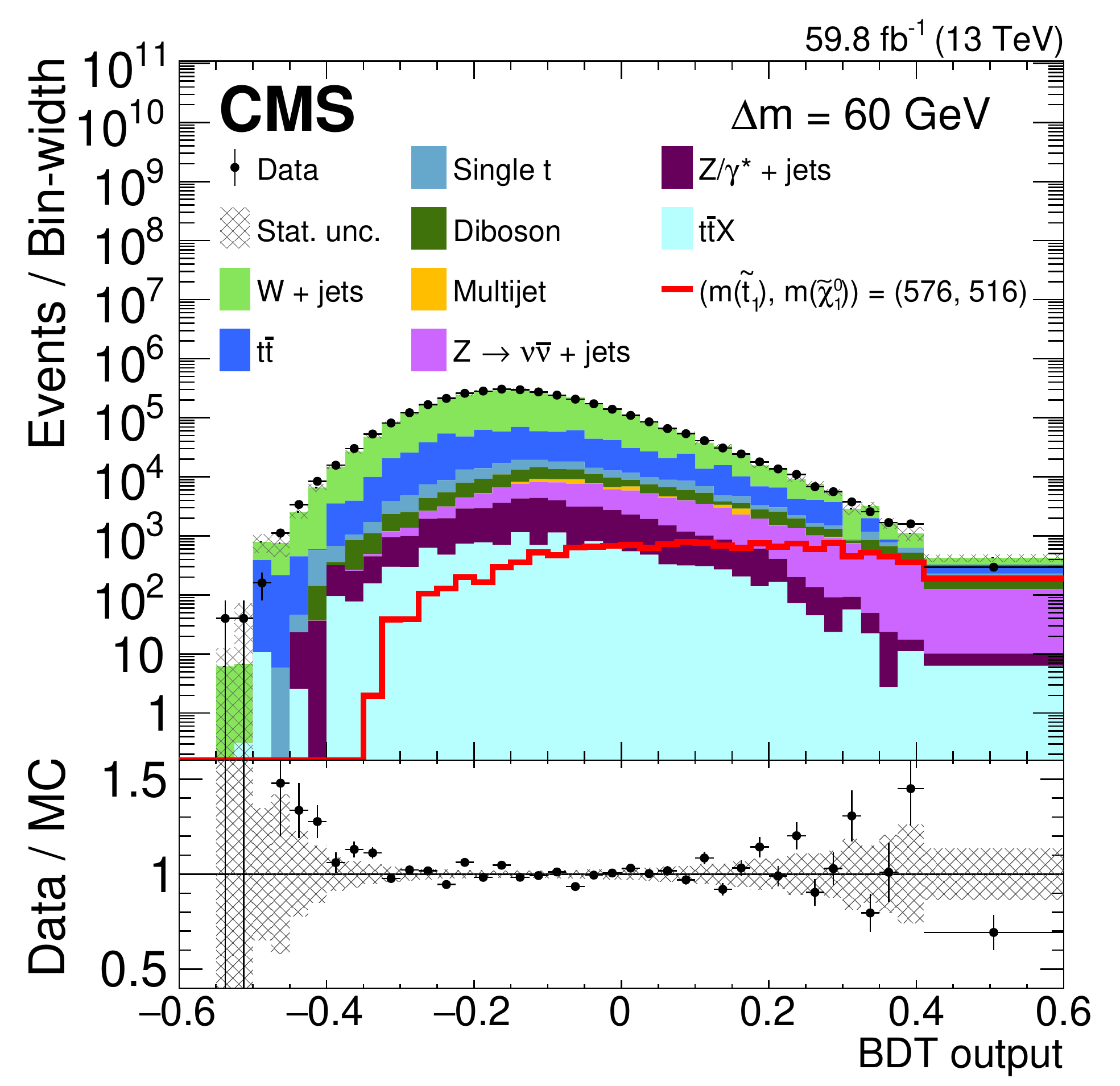} \\
\includegraphics[width=0.32\textwidth]{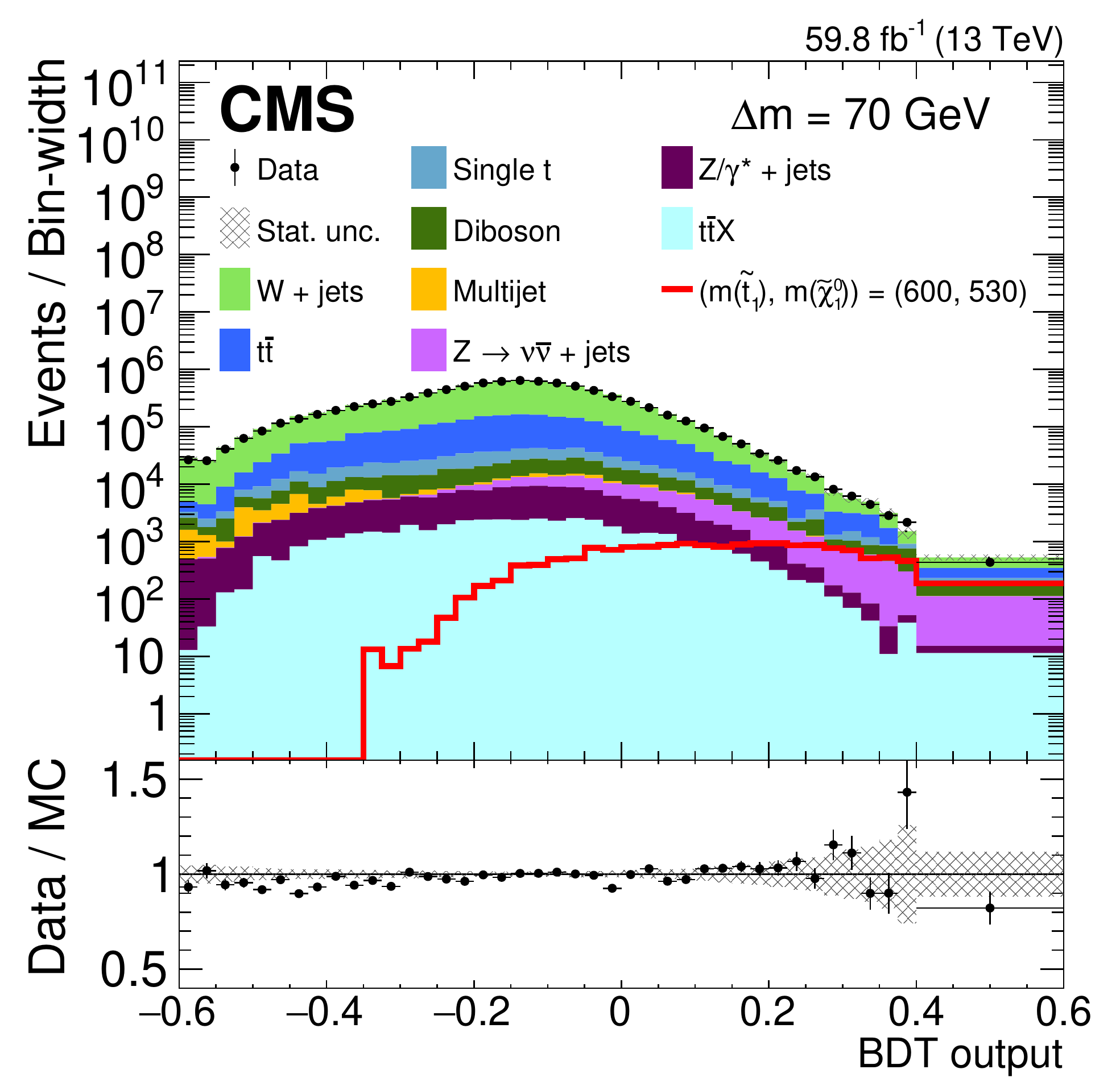} \hfil
\includegraphics[width=0.32\textwidth]{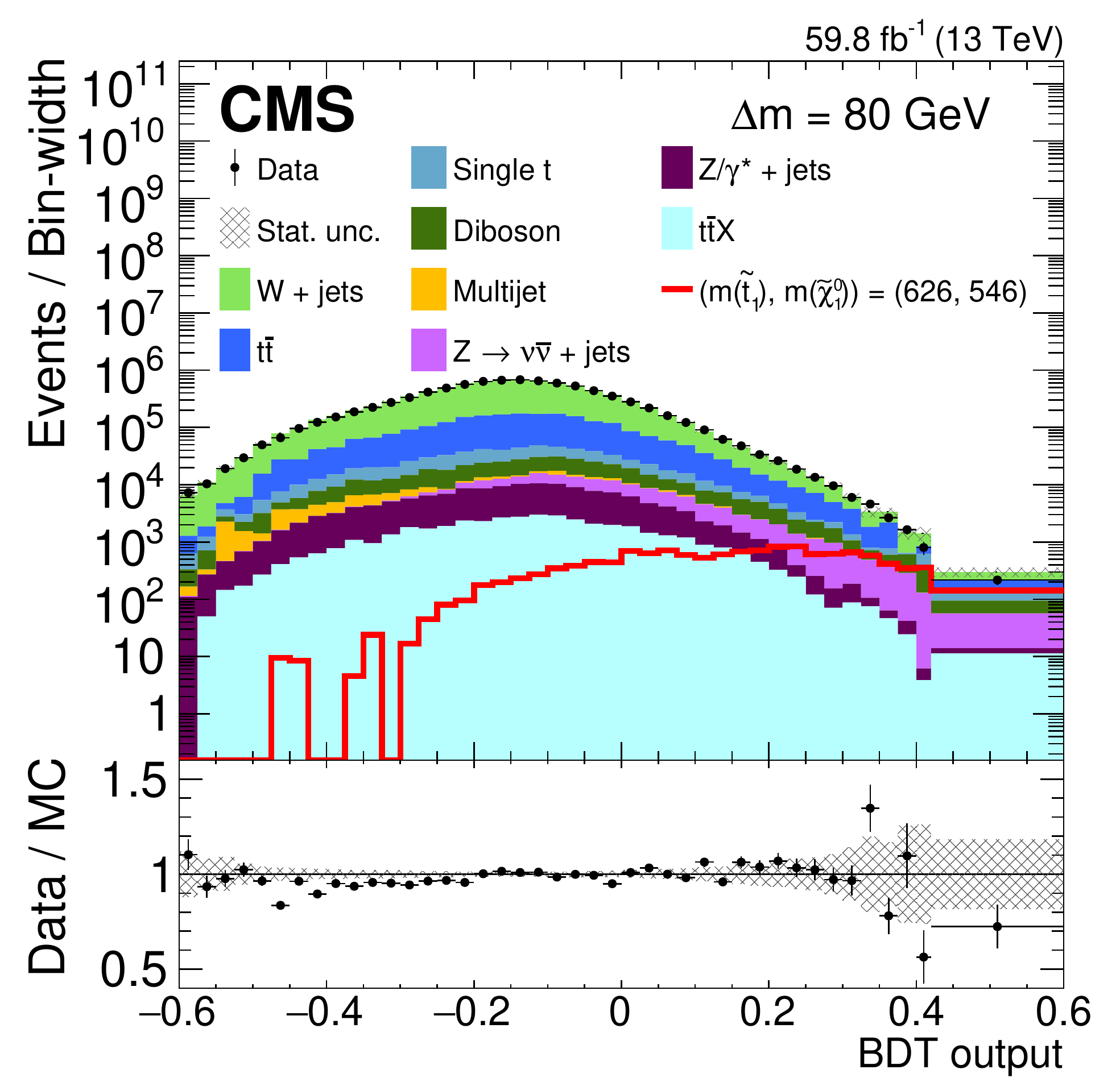} \hfil
\caption{BDT output distributions from data (points) and simulation (colored histograms) 
        after the preselection in 10\GeV steps of \dm from 10 (upper left) to 80 (lower 
        right)\GeV for the 2018 data. The last bin corresponds to the SR. 
        For each \dm value, the predicted signal distribution is shown by the
        solid red line for a representative \mmplane point, unstacked from the 
        histograms. The lower panels show the ratio of the data to the sum of the
        background predictions, with the vertical bars and shaded area giving 
        only the statistical uncertainty in the data and the simulated background, 
        respectively.
}
\label{fig:bdtdatamc2}
\end{figure}

The BDT output distributions for data and simulated SM background are shown in
Figs.~\ref{fig:bdtdatamc1} and \ref{fig:bdtdatamc2} for the 2017 and 2018 data, 
respectively. In each case a \mmplane signal point belonging to the \dm value 
for which the training has been done is also reported. 
The BDT output is found to be different for various values of \dm, which is to be expected
because of the changing mix of signal and background and the varying correlations across
the \mmplane plane, resulting in different BDT outputs 
for different \dm values. A good agreement between the data and simulation is 
observed for the BDT output distributions over the entire range, for all trainings;
the region at small BDT output values (\eg, ${<}0.3$) is dominated by background events.

\begin{figure}[htbp]
\centering
\includegraphics[width=0.45\textwidth]{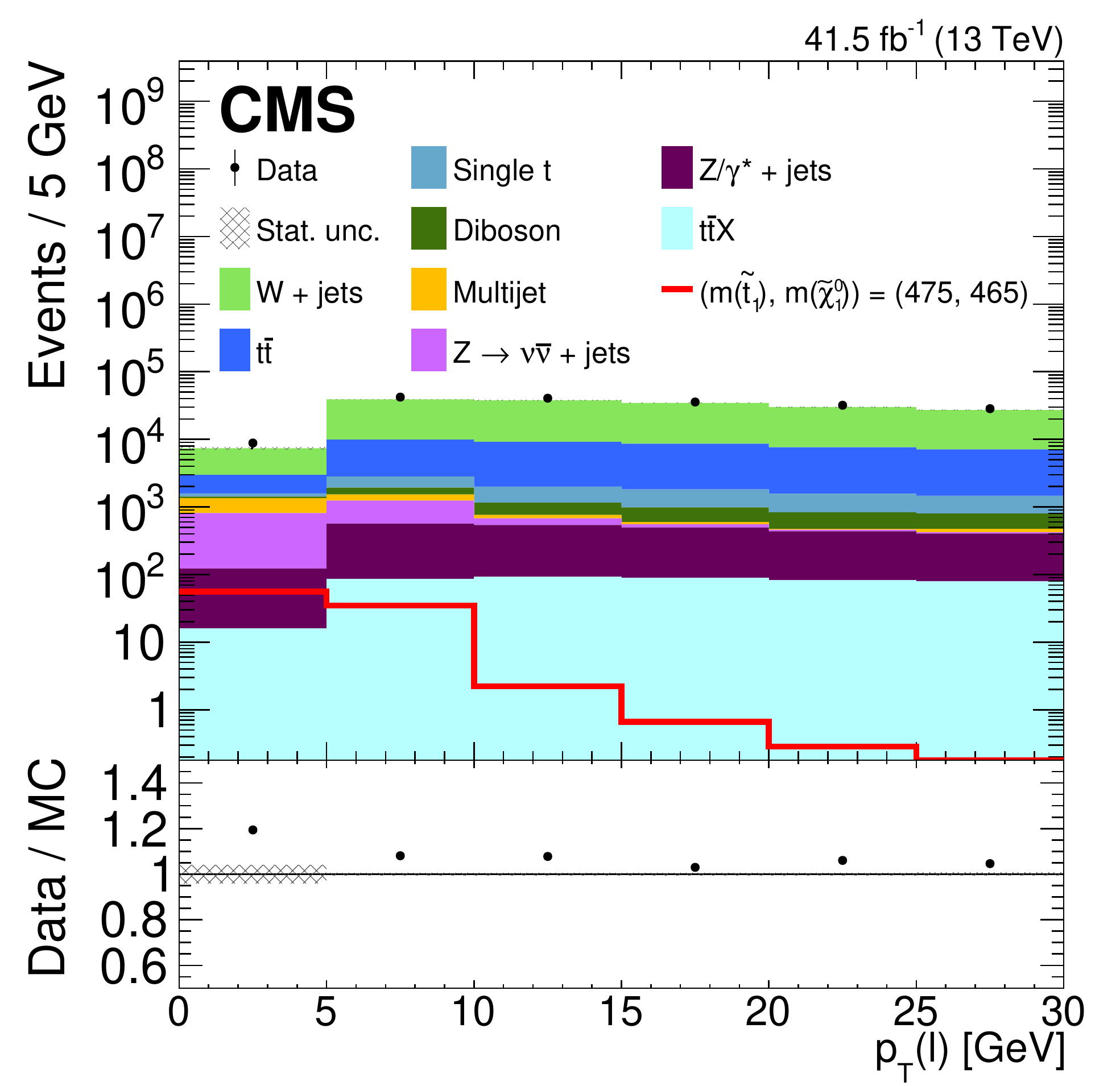} \hfil
\includegraphics[width=0.45\textwidth]{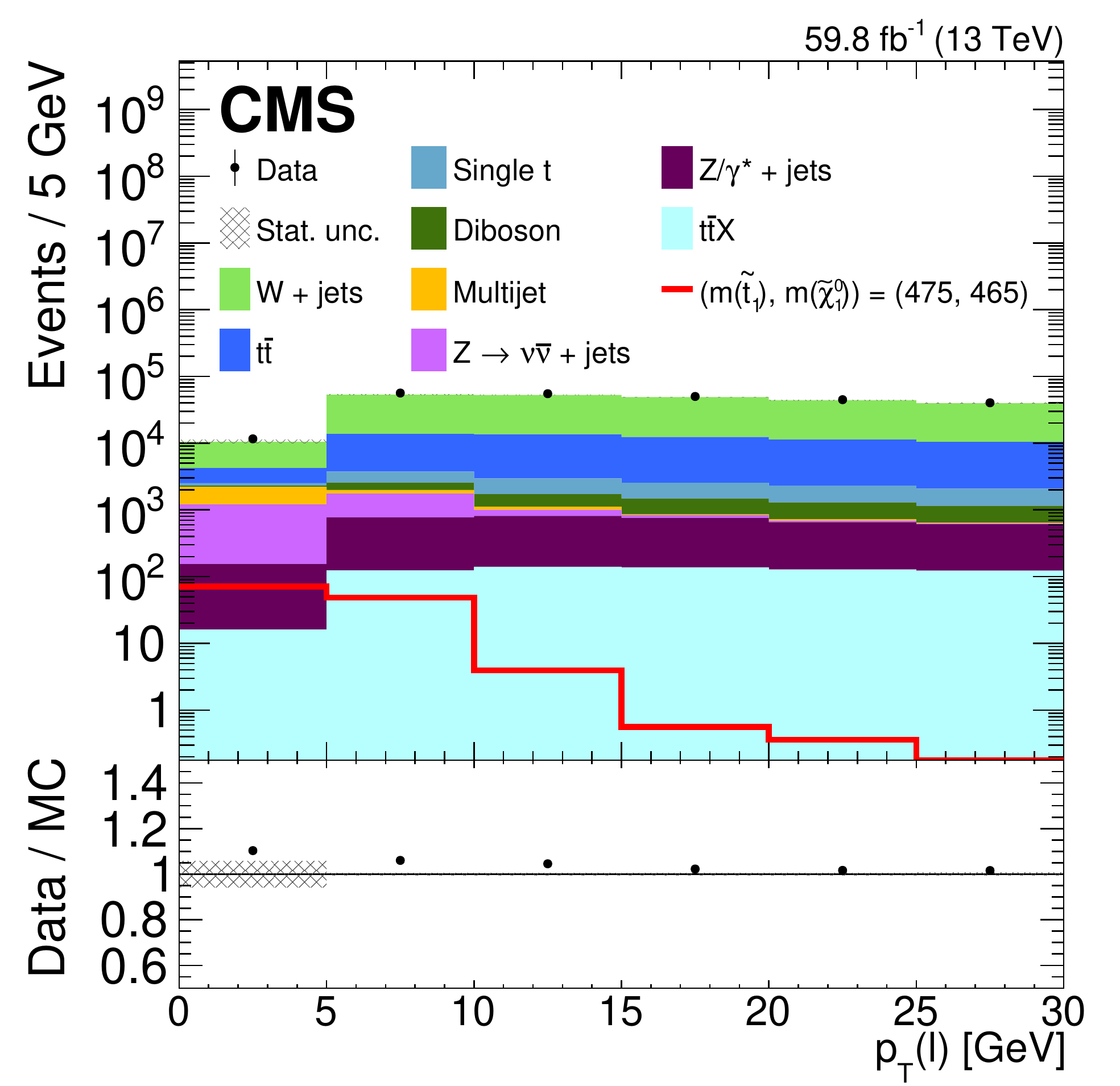} \\
\includegraphics[width=0.45\textwidth]{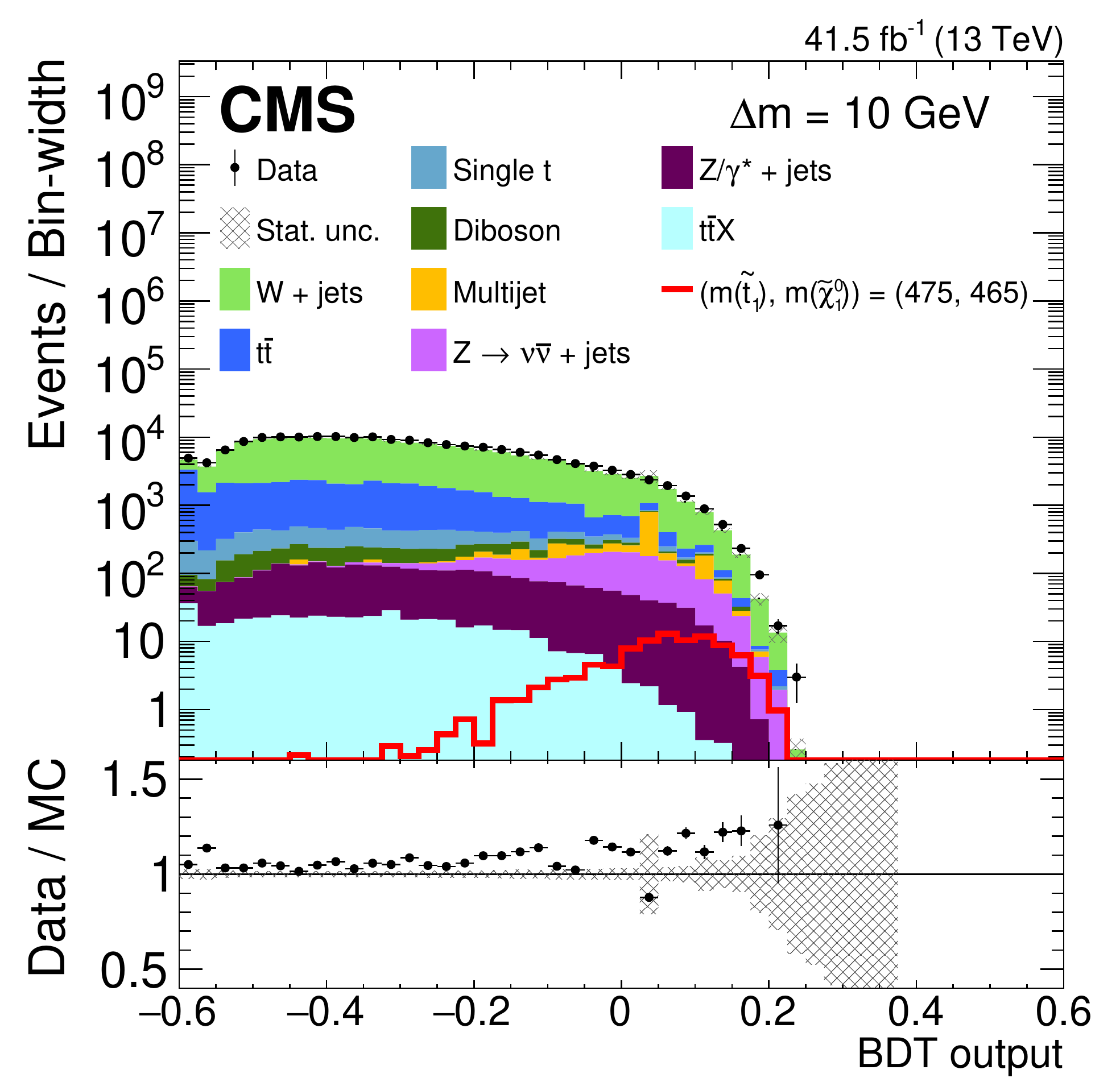} \hfil
\includegraphics[width=0.45\textwidth]{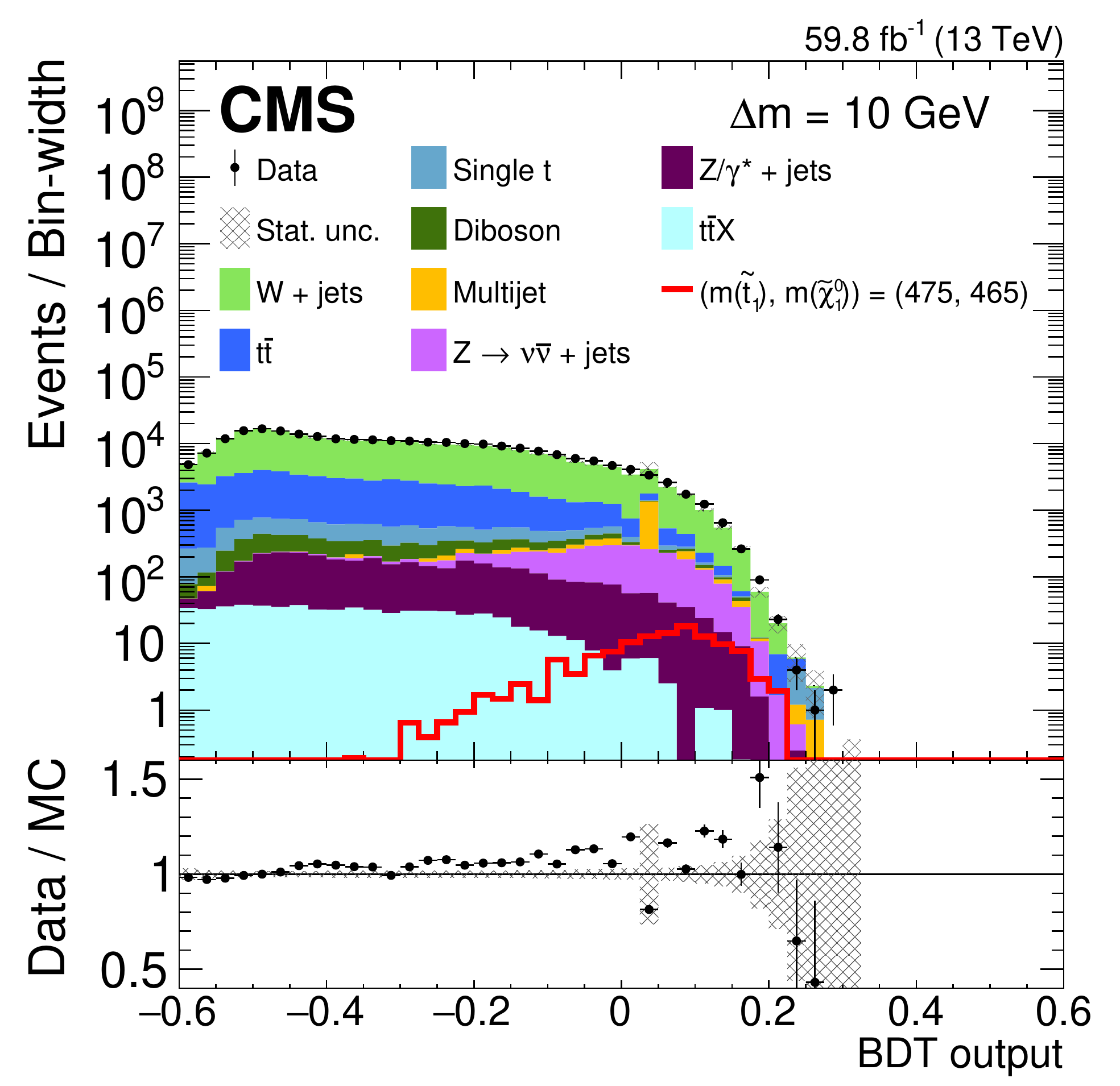} \\
\caption{
Distributions of \ptl (upper) and BDT output for $\dm=10\GeV$ (lower) in the VR 
where $200 < \ptmiss < 280\GeV$, from 2017 (left) and 2018 (right) data (points) 
and simulation (colored histograms). The predicted signal distribution is shown 
by the solid red line for \mmplane = (475, 465), unstacked from the histograms.
The lower panels show the ratio of data to the sum of the simulated SM 
backgrounds. The shaded bands indicate only the statistical uncertainty in the 
simulation predictions.
}
\label{fig:vr2}
\end{figure}

\begin{figure}[htbp]
\centering
\includegraphics[width=0.45\textwidth]{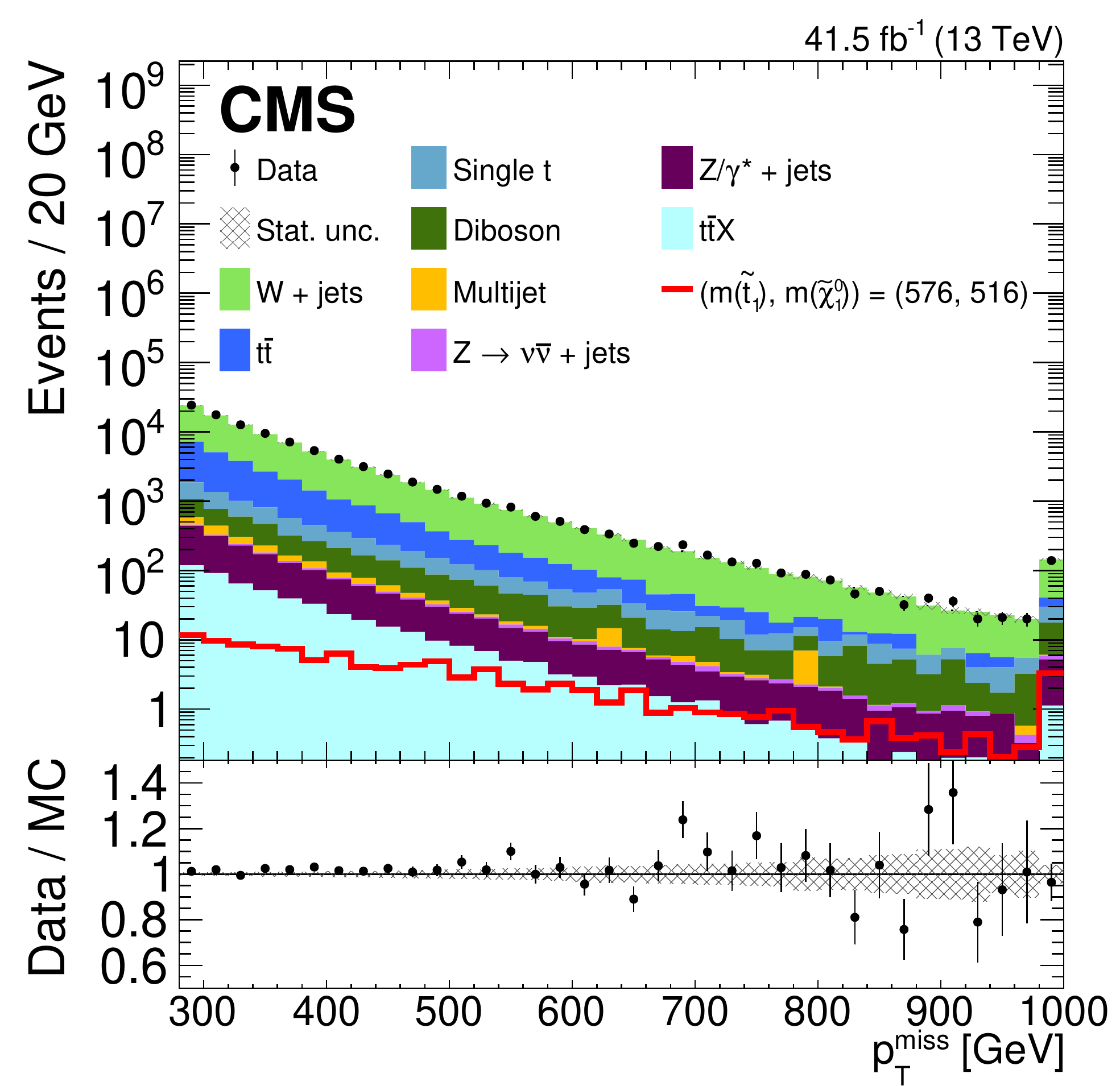} \hfil
\includegraphics[width=0.45\textwidth]{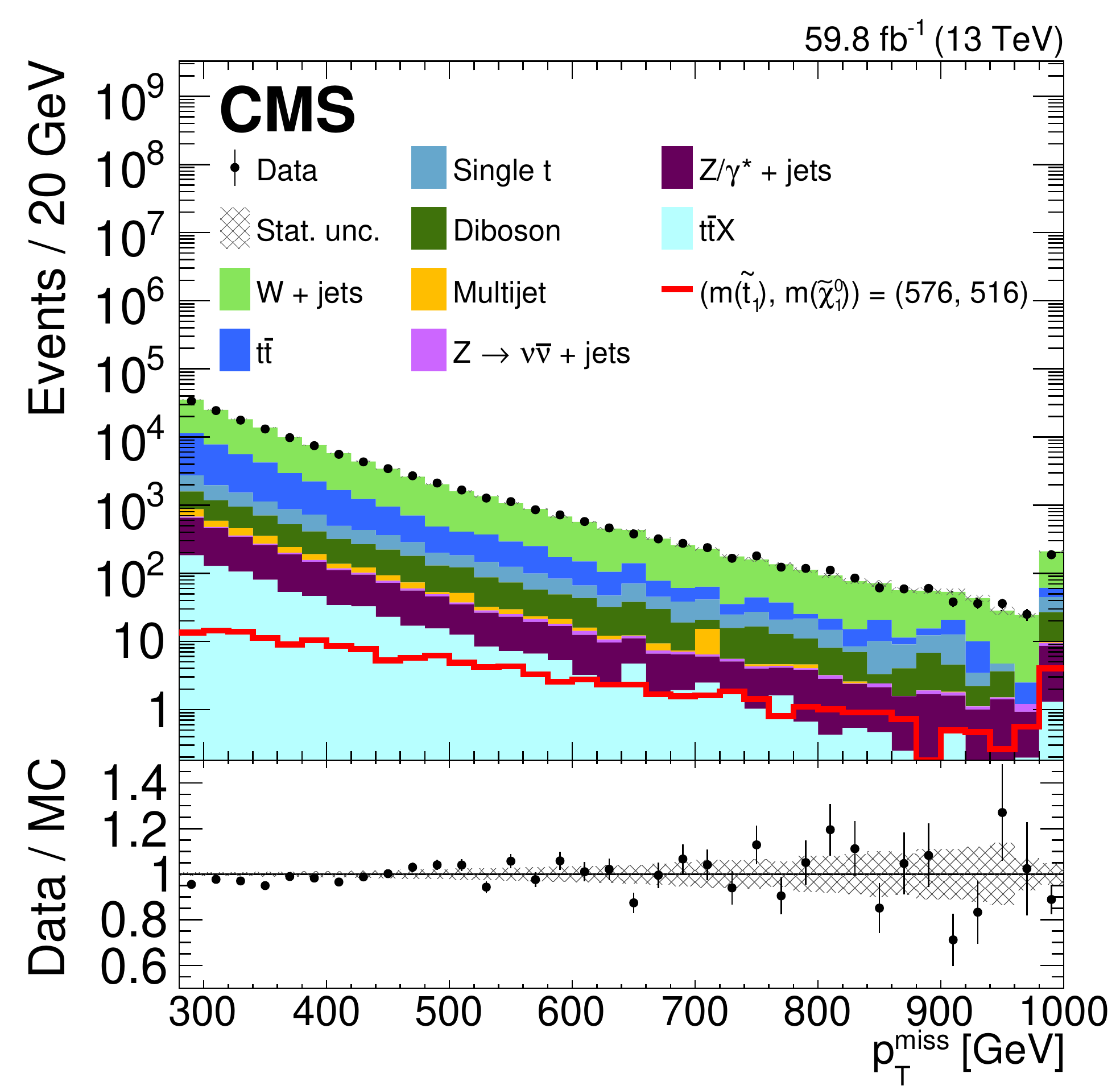} \\
\includegraphics[width=0.45\textwidth]{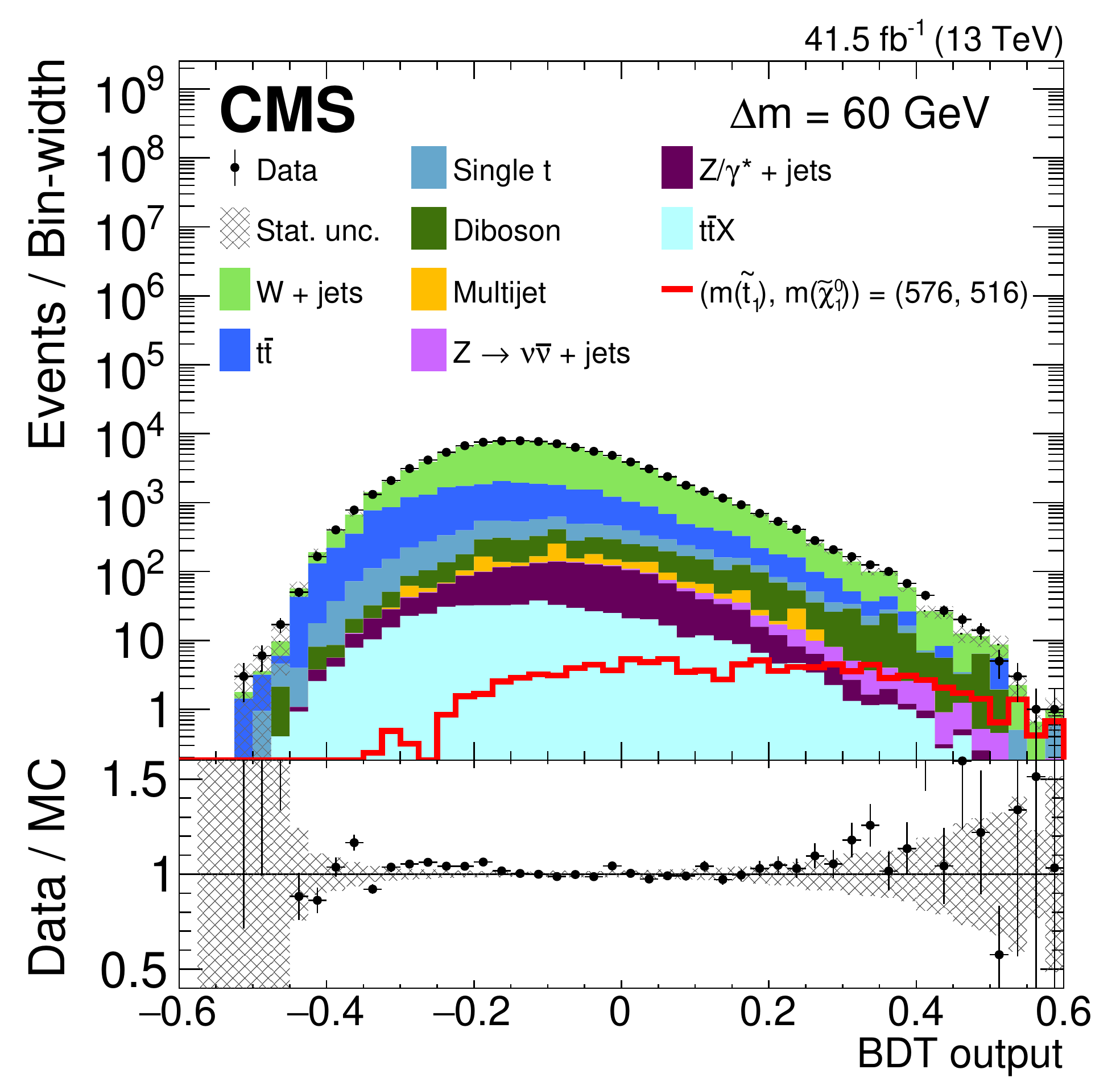} \hfil
\includegraphics[width=0.45\textwidth]{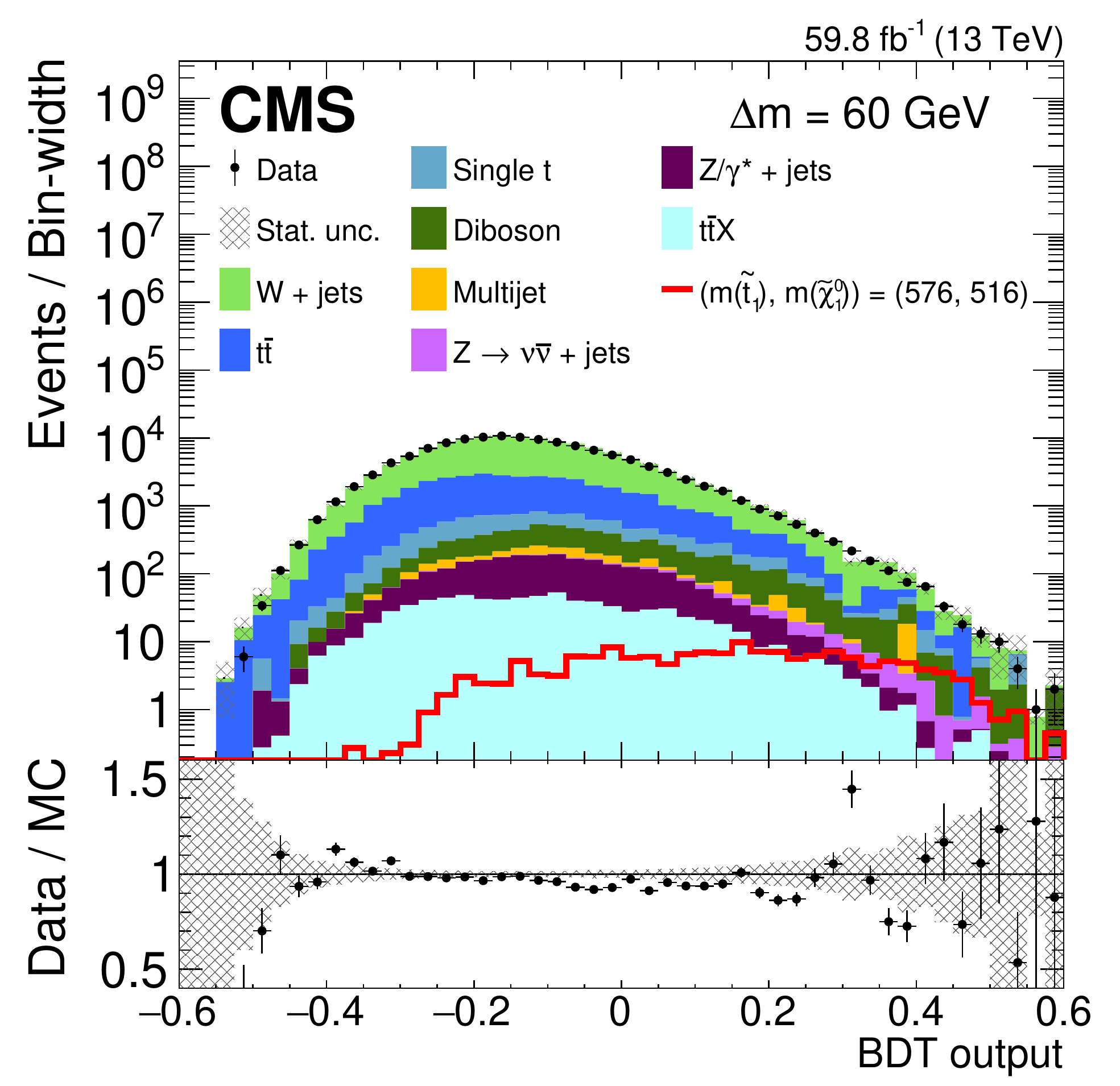} \\
\caption{
Distributions of \ptmiss (upper) and BDT output for $\dm=60\GeV$ (lower) in the 
VR where $\ptl > 30\GeV$, from 2017 (left) and 2018 (right) data (points) and 
simulation (colored histograms). The predicted signal distribution is shown 
by the solid red line for \mmplane = (576, 516), unstacked from the histograms.
The lower panels show the ratio of data to the sum of the simulated SM 
backgrounds. The shaded bands indicate only the statistical uncertainty in the 
simulation predictions.
}
\label{fig:vr3}
\end{figure}

To check the validity of the BDT output in regions depleted in signal, a set of 
validation regions (VRs) are defined. These regions are chosen to be kinematically close 
but nonoverlapping with the region selected by the preselection, while using the same 
online selection. The first VR uses the preselection requirements discussed in
Section~\ref{s:presel}, but where $200 < \ptmiss < 280\GeV$ is required. 
This VR is used to validate the BDT output for all the trained BDTs. 
The second VR also uses the preselection requirements, but where $\ptl > 30\GeV$
is required. It is used for the validation of BDTs trained for signals with $\dm < 70\GeV$.
This region is not used for BDTs trained for signals with $\dm = 70$ or $80\GeV$ 
because the entire range of the \ptl distribution is considered at preselection.
The BDT output distributions for these VRs from data are consistent with those 
from the simulation. 
Fig.~\ref{fig:vr2} illustrates for both years the \ptl distribution for the VR 
where $200 < \ptmiss < 280\GeV$, as well as the output of the BDT for $\dm=10\GeV$. 
Fig.~\ref{fig:vr3} reports for both years the \ptmiss distribution for the VR 
where $\ptl > 30\GeV$, and the output of the BDT for $\dm=60\GeV$. As observed 
in these figures, the BDT output of data is well described by the simulation. 
The mismodeling of an input variable, possibly resulting in the one of the BDT 
output, is covered by a systematic uncertainty. 
As described in Section~\ref{s:bkg}, the VRs are used to evaluate the uncertainty
in the background determination.

A signal region (SR) is defined by requiring a lower limit on each BDT output.
This limit is determined by minimizing the expected upper limit on the signal 
cross section of a benchmark \mmplane signal point at the exclusion limit of the 2016 
search. This choice implies that the benchmark signal points for the search 
from 2017 and 2018 data are at higher \stp and \lsp masses than for the 2016 
search.
The exact values of the BDT selection requirements are reported in Table~\ref{tab:mvares}.
As an illustration of the selection power of the BDT, in the case of $\dm = 80\GeV$, 
the SM background is suppressed by a factor of ${\approx}3.7\times10^3$ compared
to the preselection, while the signal is only reduced by a factor of $\approx$13.

\section{Background estimation}
\label{s:bkg}

The main background processes in this search are \wjets and \ttbar, both with a
prompt lepton, and events where the lepton arises from the decay of heavy-flavor
quarks or from misidentified hadrons that pass the lepton criteria. The latter 
category is labeled as nonprompt background. 
The processes contributing to the nonprompt background are mainly \zinv, and to 
a lesser extent, \wjets and \ttbar, where a jet is misidentified as a lepton, 
as well as the multijet background.
Furthermore, there can also be events in which a genuine lepton (mainly from 
\wjets or \ttbar) escapes detection, while a nonprompt lepton is selected. 
These three main sources of background are estimated using data, as described in
Sections~\ref{s:nonprompt} and~\ref{s:prompt}.
The background from other SM processes, such as single top quark, diboson, 
DY, and \ttX production, are estimated from simulation.
In the following section, background yields estimated using data are denoted 
by $Y$ while background yields estimated only from simulated samples are 
denoted by $N$. 

\subsection{Nonprompt background}
\label{s:nonprompt}

The nonprompt background is estimated from data using the ``tight-to-loose''
method~\cite{Khachatryan:2016kod}. 
The tight criteria correspond to the selection of the lepton as described in 
Section~\ref{s:objects}.
The loose selection is defined by 
relaxing the requirement on the isolation variable to $\Iabs<20\GeV$ for 
$\ptl<25\GeV$ and $\Irel<0.8$ for $\ptl>25\GeV$, and on the
impact parameters to $\abs{\dxy}<0.1\unit{cm}$ and $\abs{\dz}<0.5\unit{cm}$. 
A lepton passing these requirements is called a loose lepton.
The probability $\epsilon_\mathrm{TL}$ for a loose 
lepton to pass the tight criteria is measured as a function of its \pt and 
$\eta$ in a data CR that is largely dominated by multijet 
events and enriched in nonprompt leptons.
For each SR, it is defined a side-band region with the same requirements, 
but where the lepton must pass the loose criteria while failing the tight 
ones (``L!T''). The number of such events in data is denoted as $N^{L!T}$(Data). 
The number of events $N^{L!T}_\text{p}$(MC) from simulation where a vector boson
or a top quark produce a prompt lepton are subtracted from the data sample with 
a loose-not-tight lepton.
The predicted nonprompt yield $Y^\mathrm{SR}_\text{np}$ in each SR is obtained by 
weighting the resulting number of events by $\epsilon_\mathrm{TL} / (1-\epsilon_\mathrm{TL})$:
\begin{linenomath}
\begin{equation}
  Y_{\mathrm{np}}^{\mathrm{SR}} = \frac{\epsilon_{\mathrm{TL}}}{1-\epsilon_{\mathrm{TL}}} [N^{\mathrm{L!T}}
  (\text{Data}) - N_{\mathrm{p}}^{\mathrm{L!T}} (\text{MC})].
\label{eq:fakes}
\end{equation}
\end{linenomath}

\subsection{Dominant prompt backgrounds}
\label{s:prompt}

To estimate the prompt contributions from the \wjets and \ttbar processes, a method based 
on the number of these background events observed in data CRs is used. The method uses the output of the 
BDT, and a transfer factor between the CR and the SR, obtained from simulation. 
This factor, of the order $10^{-3}$ for both backgrounds and for both years, is
the ratio of the number of predicted events in the SR, $N^\mathrm{SR}_\text{p}$,
to the one in the CR, $N^\mathrm{CR}_\text{p}$. The estimated yield 
$Y^\mathrm{SR}_\text{p}$ of the dominant prompt background in the SR, estimated 
independently per process and per year, is then determined using:
\begin{linenomath}
\begin{equation}
\begin{aligned}
  Y^\mathrm{SR}_\text{p}(X) = \frac{N^\mathrm{SR}_\text{p}(X)}
  {N^\mathrm{CR}_\text{p}(X)} \left[N^\mathrm{CR}(\text{Data}) - 
  N^\mathrm{CR}_\text{p}(\text{non-}X) - Y^\mathrm{CR}_\text{np}\right],
\label{eq:prompt}
\end{aligned}
\end{equation}
\end{linenomath}
where $X$ refers to the background process being estimated, either \wjets or 
\ttbar, and where the terms prompt and nonprompt refer to their definition as 
given at the beginning of Section~\ref{s:bkg}.
To obtain a data sample enriched in the backgrounds being estimated, a CR is defined
by applying the preselection criteria, with the additional requirement $\mathrm{BDT} < 0$. The 
number of such events is denoted as $N^\mathrm{CR}(\text{Data})$. To enrich 
the CR in \wjets or \ttbar events, the number of loosely \PQb tagged jets is 
required to be zero, or the number of tightly \PQb tagged jets to be at least one, respectively,
where loose and tight were discussed in Section~\ref{s:objects} \cite{CMS-DP-2018-058}.
The purity of \wjets and \ttbar processes in the corresponding CRs is approximately 93\% 
and 78\%, respectively. The level of signal contamination in the CR is well below 5\%.
The number $N^\mathrm{CR}_\text{p}(\text{non-}X)$ is the number of prompt background
events other than the process being estimated in the CR, estimated from simulation 
and subtracted from the number of data events; \eg: if $X =$ \wjets, this term 
includes \ttbar, and vice versa. The yield $Y^\mathrm{CR}_\text{np}$, which is 
the predicted number of nonprompt background in the CR, is also subtracted.

\section{Summary of systematic uncertainties}
\label{s:sys}

Processes for which the absolute yield is predicted by simulation are
subject to systematic uncertainties in the determination of the integrated
luminosity, which is estimated year-by-year with uncertainties in the 1.2--2.5\% range
~\cite{CMS-PAS-LUM-17-004,CMS-PAS-LUM-18-002}. All simulated 
samples are subject to experimental uncertainties in the jet energy scale (JES) 
and jet energy resolution (JER). The uncertainties arising from miscalibration of
the JES are estimated by varying the jet energy corrections up and down 
by one standard deviation of their uncertainties and propagating the effect to the calculation 
of \ptmiss. Differences in the JER between data and simulation 
are accounted for by smearing the momenta of jets in simulation. The
uncertainties corresponding to the \PQb tagging efficiencies and
misidentification rates for tagging light-flavored quark or gluon jets as
\PQb jets have been evaluated for all simulated samples. The systematic
uncertainties in the scale factors applied to the simulated samples for trigger 
and lepton efficiencies are taken into account. 
The uncertainty due to the simulation of pileup for simulated background
processes is estimated by varying the inelastic $\Pp\Pp$ cross section by 
4.6\%~\cite{Sirunyan:2018nqx}. An uncertainty of
50\% is assigned to the cross sections of all backgrounds whose yields are 
predicted from simulation.

The estimation of nonprompt backgrounds, as described in Section~\ref{s:nonprompt}, 
depends on the tight-to-loose fraction $\epsilon_\mathrm{TL}$, which is sensitive 
to the flavor content of jets. The systematic uncertainty arising from this source
in the measurement region is estimated by changing the \PQb tagging requirement in 
the \PQb veto to demand at least one \PQb tagged jet using the medium working point.
The resulting uncertainty ranges from 3 to 90\% from low to high lepton \pt, 
respectively. The method is also tested by repeating this procedure on the simulated
event samples, where any variations in the background determination are considered
as systematic uncertainties and added in quadrature to the aforementioned uncertainty.

The systematic uncertainties associated with the predictions of \wjets and 
\ttbar processes are based on the comparison of two methods: one to assess the 
closure of the background prediction method, and the other to evaluate the 
effect of the modeling of the BDT output distribution. The closure method 
measures differences between the predicted number of events (obtained from 
Eq.~(\ref{eq:prompt})) and the observed number of data events in both VRs, as 
defined in Section~\ref{s:sel}, where the statistical uncertainty in the number 
of CR events is taken into account. 
Uncertainties in modeling the BDT output distribution, which can affect the 
background prediction, are assessed by comparing the ratio of the BDT output 
distributions for data to the background prediction in the CR with the ratio 
in the SR, for the two VRs.
To be conservative, the uncertainties are evaluated in the two VRs for both 
methods, and the largest value is used.
These uncertainties range from 10 to 20\% for the prediction of \wjets events and
from 8 to 80\% for the estimation of \ttbar processes over the various SRs. 
The estimations of the \wjets and \ttbar backgrounds rely partially on the simulation
and are therefore sensitive to theoretical uncertainties in the modeling of ISR.
For the \ttbar process, half of the ISR correction is assigned as the systematic
uncertainty, which also applies to the simulated signal samples.
For the \wjets process, the difference between the ISR-corrected and uncorrected 
simulation is taken as the systematic uncertainty.

Uncertainties from unknown higher-order theoretical effects are estimated through 
uncorrelated variations of the renormalization and factorization scales by factors 
of 0.5, 1, and 2~\cite{syscalc}. Finally, differences between the fast and the full 
\GEANTfour-based modelings of \ptmiss are used as the corresponding systematic 
uncertainty and assigned to the signal yields. 
The statistical uncertainty in the signal simulation samples of 3 to 20\% over the 
various SRs is included as a systematic uncertainty.
The relative systematic uncertainties in the signal from the various sources, and 
the total relative systematic uncertainties in the \wjets, \ttbar, and nonprompt
backgrounds, are given in Table~\ref{tab:mvasys} as ranges over the eight SRs.

\begin{table}[!tbp]
\centering
\topcaption{The relative systematic uncertainties in percent from the different 
    sources in the signal and the total relative uncertainty in the \wjets, \ttbar, and 
    nonprompt background predictions, shown separately for the 2017 and 
    2018 data analysis. The ranges given are across the eight SRs.
    The ``\NA'' symbol means that a given source of uncertainty is not applicable. 
}
\begin{tabular}{lcccc}
                        & 
\multicolumn{2}{c}{2017} & 
\multicolumn{2}{c}{2018} \\
\hline
Source                            & Background & Signal & Background & Signal \\
\hline
Integrated luminosity             & \NA        & 2.3    & \NA        & 2.5    \\
JES                               & 0--2       & 3--9   & 0--2       & 5--10  \\
JER                               & 0--1       & 0--1   & 0--1       & 0--1   \\
\PQb tagging                     & 0--1       & 0--6   & 0--1       & 0--1   \\
Trigger                           & 0--1       & 1      & 0--1       & 1      \\
Lepton efficiency                 & 0--1       & 0--1   & 0--1       & 0--1   \\
Pileup                            & 1--5       & 0--3   & 1--4       & 0--1   \\
ISR (\ttbar and signal)           & 0--1       & 0--5   & 0--1       & 0--5   \\
ISR (\wjets)                      & 0--4       & \NA    & 0--4       & \NA    \\
Renorm./Fact. scales              & 0--7       & 0--1   & 0--10      & 0--1   \\
\ptmiss modeling (\textsc{FastSim}) & \NA        & 0--2   & \NA        & 0--2   \\
\wjets total                      & 2--6       & \NA    & 4--9       & \NA    \\
\ttbar total                      & 1--5       & \NA    & 2--7       & \NA    \\
Nonprompt lepton total            & 2--5       & \NA    & 2--4       & \NA    \\
\end{tabular}
\label{tab:mvasys}
\end{table}

To combine the results from the different data-taking years, systematic 
uncertainties whose sources are exactly the same for the different years are taken
as fully correlated.
This includes the uncertainty in the theoretical cross sections, pileup, JES,
the reweighting of the \wjets sample, the renormalization and factorization
scales, and the prediction of the \wjets, \ttbar, and nonprompt backgrounds.
The systematic uncertainty in the integrated luminosity has multiple 
components and is thus considered as partially correlated between the 
years~\cite{CMS-LUM-17-003,CMS-PAS-LUM-17-004,CMS-PAS-LUM-18-002}, as is the 
systematic uncertainty in the \PQb tagging procedure.

\section{Results and interpretation}
\label{s:res}

\begin{figure}[!htbp]
\centering
\includegraphics[width=0.85\textwidth]{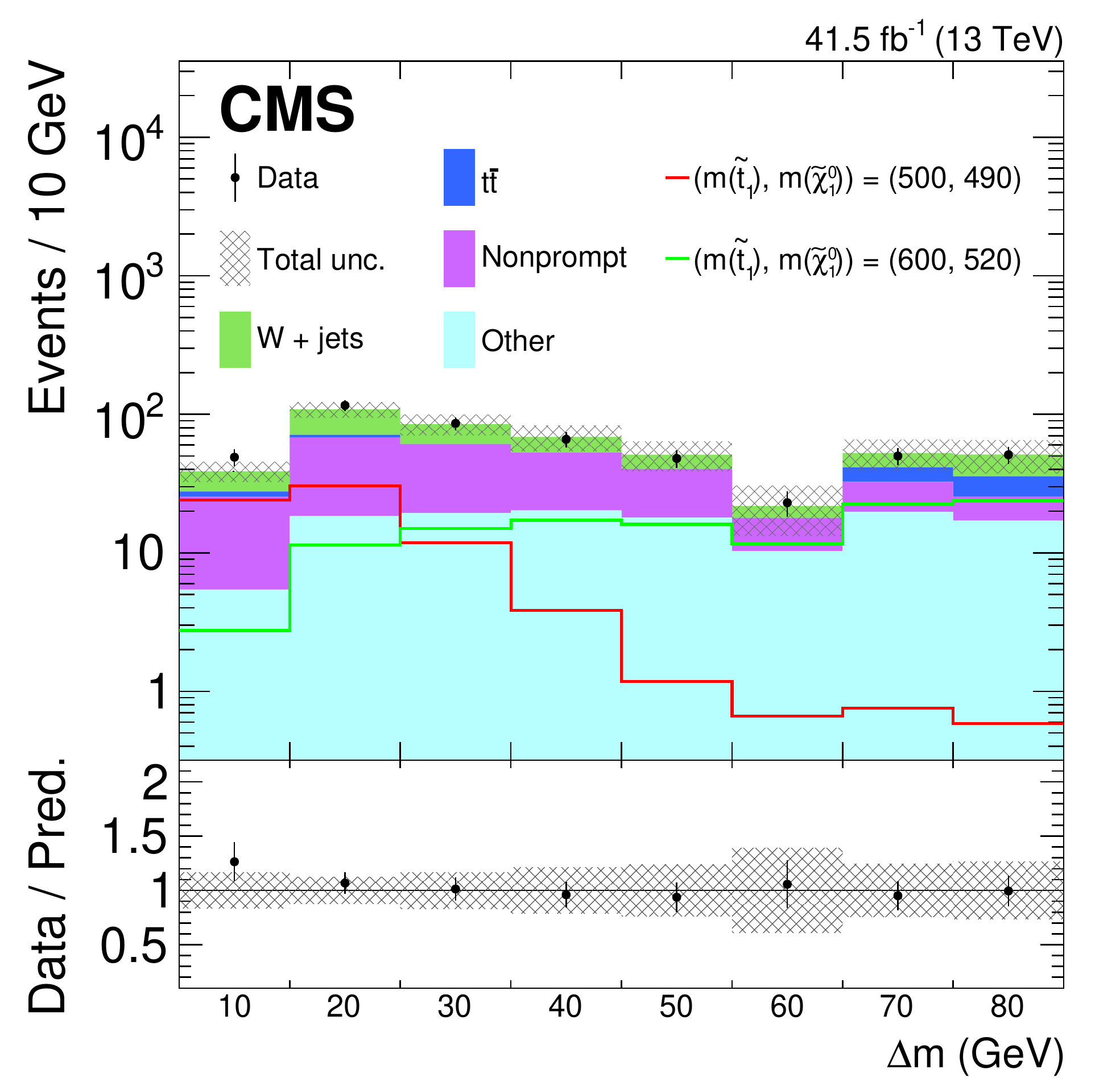}
\caption{The observed yields in data (points) and the predicted background 
components (colored histograms) in the eight SRs for the 2017 data. The 
vertical bars on the points give the statistical uncertainty in the data.
The hatched area shows the total uncertainty in the sum of the backgrounds.
The expected yields for two signal points with \mmplane = (500, 490) and 
(600, 520)\GeV are also given by the lines, unstacked from the histograms. 
The lower panel shows the ratio of the number of observed events to the predicted 
total background.  The vertical bars on the points give the statistical 
uncertainty in the ratio and the hatched area the total uncertainty.
}
\label{fig:SR2017}
\end{figure}

\begin{figure}[!htbp]
\centering
\includegraphics[width=0.85\textwidth]{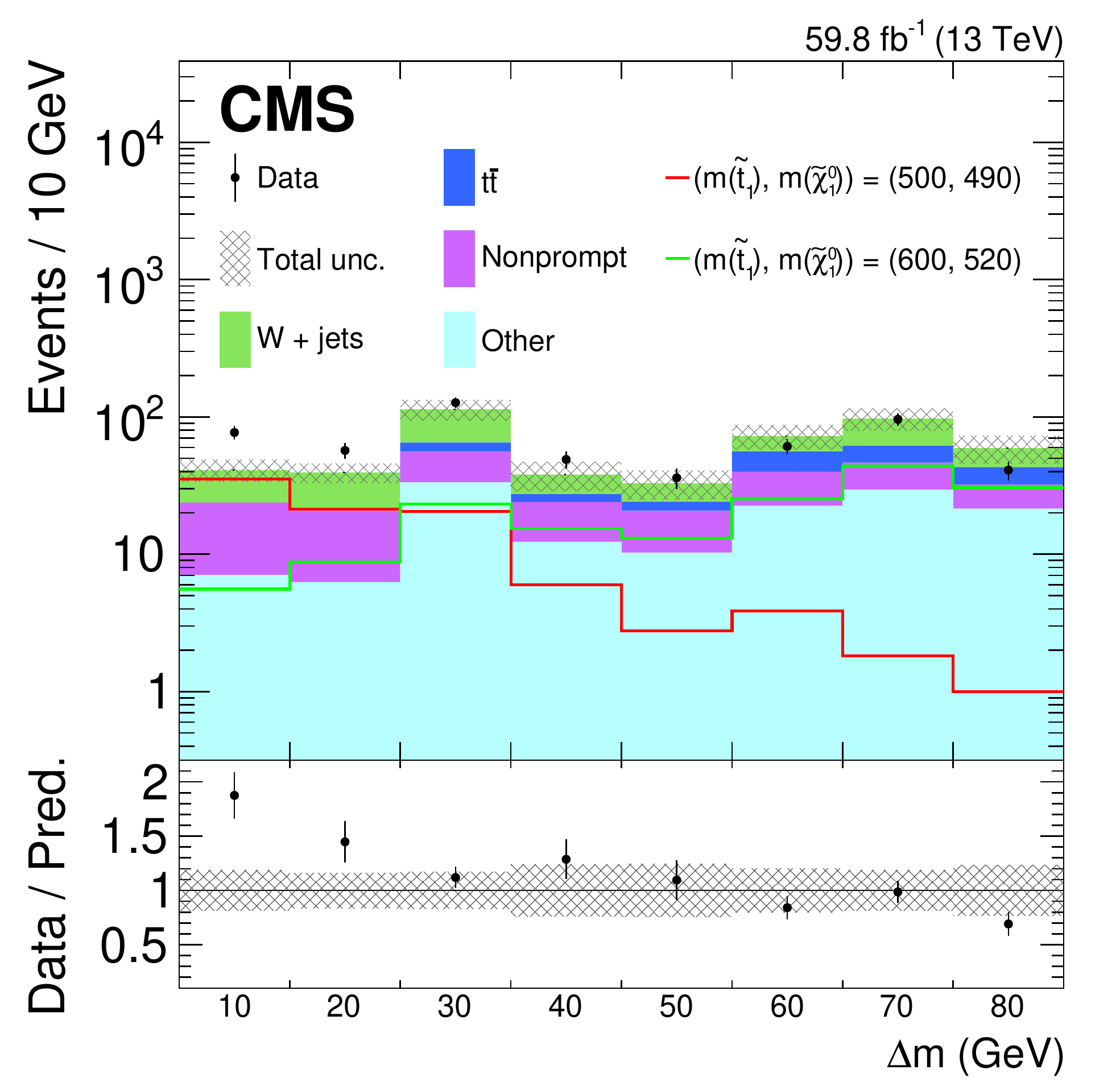}
\caption{The observed yields in data (points) and the predicted background 
components (colored histograms) in the eight SRs for the 2018 data. The 
vertical bars on the points give the statistical uncertainty in the data.
The hatched area shows the total uncertainty in the sum of the backgrounds.
The expected yields for two signal points with \mmplane = (500, 490) and 
(600, 520)\GeV are also given by the lines, unstacked from the histograms. 
The lower panel shows the ratio of the number of observed events to the predicted 
total background.  The vertical bars on the points give the statistical 
uncertainty in the ratio and the hatched area the total uncertainty.
}
\label{fig:SR2018}
\end{figure}

The observed and expected numbers of signal and background events from the 2017-18 
data analysis for the eight values of \dm are given in Table~\ref{tab:mvares} and 
shown in Figs.~\ref{fig:SR2017} and \ref{fig:SR2018}. 
The predictions and the associated uncertainties in these figures are given before a 
profiled likelihood fit~\cite{frequentist_limit,Read:2002hq,LHC-HCG} is performed. 
The post-fit uncertainties do not get reduced because of the lack of constraints from 
a single bin.
It should be noted that the background composition varies for the same \dm 
region for different years. This is because an independent BDT is trained per 
\dm and per year, with a different selection on its output.
There is good agreement between the observed and predicted numbers of events for all 
SRs. The largest difference is for $\dm = 10\GeV$, where there are 1.1 and 2.9 standard 
deviations (local significance) excesses of data events over the predicted background 
for the 2017 and 2018 data, respectively. The 2016 analysis had a similar excess for the 
same \dm value, corresponding to 0.7 standard deviations. None of these excesses is 
statistically significant, so it is concluded that there is no evidence for 
direct top squark production.

\begin{table}[!htbp]
\centering
\topcaption{The predicted number of \wjets, \ttbar, nonprompt, and other ($N^\mathrm{SR}$(Other)) 
background events and their sum ($N^\mathrm{SR}$(Total)), in the eight SRs for the 2017 and 2018 data analysis.  
The first 3 predicted yields are derived from data, while the yields of the other 
background processes come from simulation. The uncertainties shown are the quadratic sum 
of the statistical and systematic uncertainties given in Table~\ref{tab:mvasys} 
for all the background processes.
The corresponding \dm and BDT output threshold values for each SR are displayed in the 
first and second columns, respectively, and the observed number of events in data is 
shown in the last column.}
\resizebox{\columnwidth}{!}{
\begin{tabular}{ccc r@{ }c@{ }l r@{ }c@{ }l r@{ }c@{ }l r@{ }c@{ }l r@{ }c@{ }l c}
Year & 
\dm\ ({\GeVns}) &
BDT $>$ & 
\multicolumn{3}{c}{$Y^\mathrm{SR}_\text{p}$(\wjets)} &
\multicolumn{3}{c}{$Y^\mathrm{SR}_\text{p}$(\ttbar)} &
\multicolumn{3}{c}{$Y^\mathrm{SR}_\text{np}$} & 
\multicolumn{3}{c}{$N^\mathrm{SR}$(Other)} & 
\multicolumn{3}{c}{$N^\mathrm{SR}\text{(Total)}$} & 
$N^\mathrm{SR}\text{(Observed)}$ \\
\hline
\multirow{8}{*}{2017}
 & 10 & 0.31 & 11.0 &$\pm$&  2.4 &  2.2 &$\pm$&  2.9 & 20.1 &$\pm$& 3.5 &  5.4 &$\pm$&  3.7 &  38.8 &$\pm$&  6.3 &  49 \\
 & 20 & 0.32 & 37.4 &$\pm$&  4.6 &  3.3 &$\pm$&  5.2 & 49.6 &$\pm$& 7.0 & 18.4 &$\pm$&  9.3 & 109   &$\pm$& 14   & 116 \\
 & 30 & 0.38 & 23.8 &$\pm$&  3.8 &  0.0 &$\pm$&  7.2 & 41.7 &$\pm$& 6.1 & 19.4 &$\pm$&  9.9 &  85   &$\pm$& 14   &  86 \\
 & 40 & 0.40 & 15.9 &$\pm$&  2.6 &  0.0 &$\pm$&  8.1 & 32.6 &$\pm$& 5.5 & 20   &$\pm$& 10   &  69   &$\pm$& 15   &  66 \\
 & 50 & 0.43 & 10.9 &$\pm$&  2.0 &  0.0 &$\pm$&  6.7 & 22.3 &$\pm$& 4.0 & 17.9 &$\pm$&  9.2 &  51   &$\pm$& 12   &  48 \\
 & 60 & 0.47 &  3.9 &$\pm$&  0.8 &  0.0 &$\pm$&  6.2 &  7.6 &$\pm$& 2.2 & 10.3 &$\pm$&  5.4 &  21.8 &$\pm$&  8.5 &  23 \\
 & 70 & 0.39 & 11.1 &$\pm$&  2.0 &  8.9 &$\pm$&  7.6 & 12.9 &$\pm$& 2.9 & 19.7 &$\pm$&  9.8 &  53   &$\pm$& 13   &  50 \\
 & 80 & 0.41 & 15.6 &$\pm$&  4.3 & 10.3 &$\pm$&  9.7 &  8.3 &$\pm$& 2.2 & 17.1 &$\pm$&  8.2 &  51   &$\pm$& 14   &  51 \\
[\cmsTabSkip]
\multirow{8}{*}{2018}
 & 10 & 0.32 & 17.3 &$\pm$&  4.3 &  0.0 &$\pm$&  2.4 & 16.7 &$\pm$& 3.6 &  7.1 &$\pm$&  4.5 &  41.1 &$\pm$&  7.6 &  77 \\
 & 20 & 0.39 & 18.4 &$\pm$&  2.8 &  0.3 &$\pm$&  3.1 & 14.5 &$\pm$& 3.4 &  6.3 &$\pm$&  3.5 &  39.4 &$\pm$&  6.4 &  57 \\
 & 30 & 0.35 & 48.5 &$\pm$&  8.1 &  9.1 &$\pm$&  9.4 & 22.5 &$\pm$& 4.8 & 33   &$\pm$& 14   & 114   &$\pm$& 19   & 127 \\
 & 40 & 0.43 & 10.7 &$\pm$&  3.1 &  3.4 &$\pm$&  4.5 & 11.7 &$\pm$& 2.9 & 12.3 &$\pm$&  6.7 &  38.1 &$\pm$&  9.1 &  49 \\
 & 50 & 0.46 &  8.7 &$\pm$&  3.0 &  3.4 &$\pm$&  4.5 & 10.5 &$\pm$& 2.8 & 10.3 &$\pm$&  5.2 &  32.9 &$\pm$&  8.0 &  36 \\
 & 60 & 0.41 & 16.5 &$\pm$&  4.7 & 16.2 &$\pm$&  8.8 & 17.3 &$\pm$& 3.8 & 22   &$\pm$& 10   &  72   &$\pm$& 15   &  61 \\
 & 70 & 0.40 & 35.6 &$\pm$&  8.7 & 15.2 &$\pm$&  8.6 & 16.9 &$\pm$& 5.2 & 30   &$\pm$& 12   &  97   &$\pm$& 18   &  96 \\
 & 80 & 0.42 & 16.3 &$\pm$&  3.7 & 10.9 &$\pm$&  7.8 & 10.7 &$\pm$& 4.3 & 21.5 &$\pm$&  9.8 &  59   &$\pm$& 14   &  41 \\
\end{tabular}
}
\label{tab:mvares}
\end{table}

\begin{figure}[!htbp]
\centering
\includegraphics[width=0.85\textwidth]{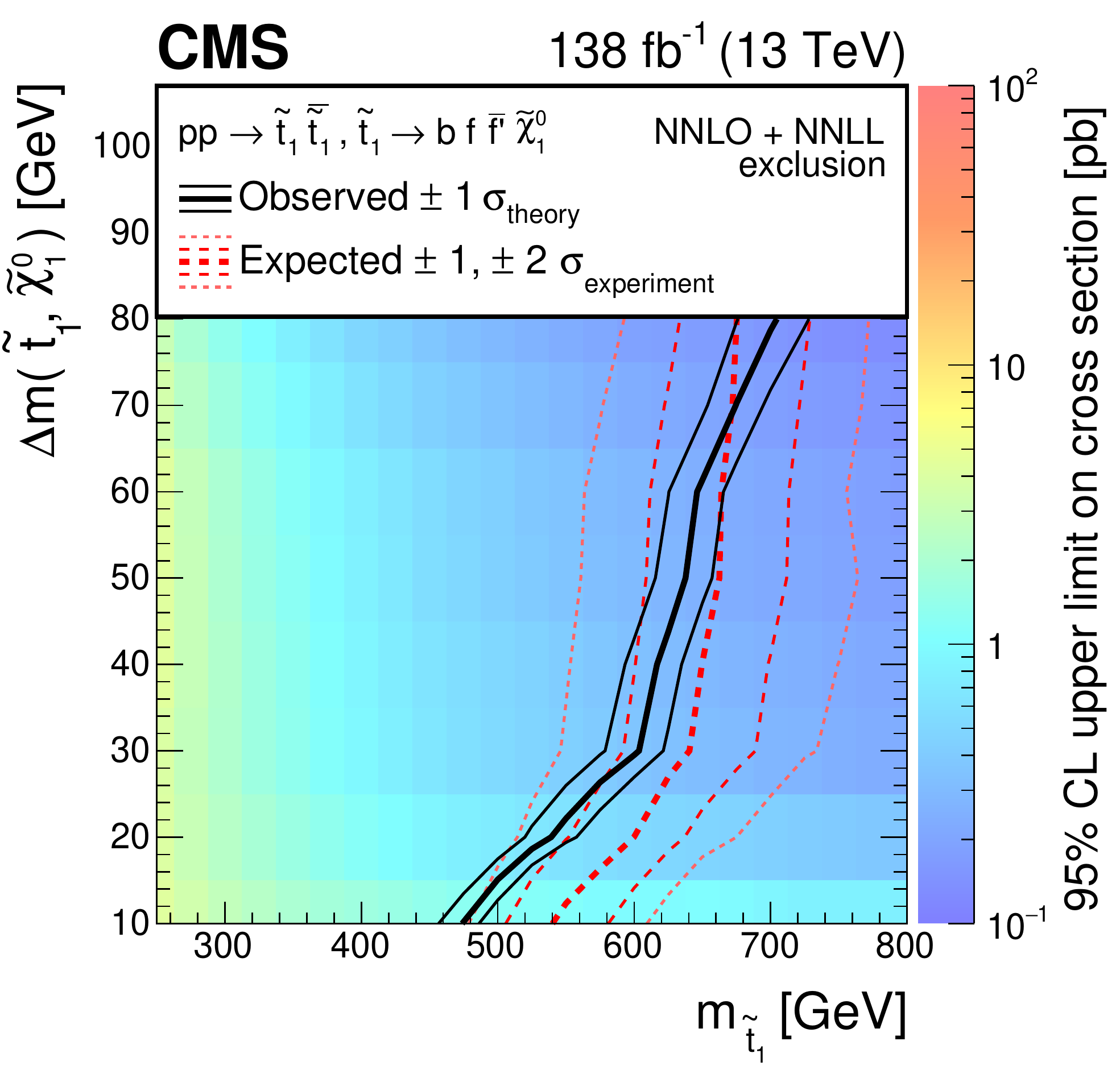}
\caption{The 95\% \CL upper limits in the (m(\stp), \dm) plane on the cross
section for the production and four-body decay of the top squark using the combined 2016,
2017, and 2018 data. The color shading represents the observed upper limit for a given
point in the plane, using the color scale to the right of the figure. The solid black and
dashed red lines show the observed and expected 95\% \CL lower limits, respectively, on
m(\stp) as a function of \dm. The thick lines give the central values of the limits. 
The corresponding thin lines represent the $\pm$ 1 standard deviation ($\sigma_{\text{theory}}$) 
variations in the limits due to the theoretical uncertainties in the case of the observed
limits, and $\pm$ 1 and 2 standard deviation ($\sigma_{\text{experiment}}$) variations due to the
experimental uncertainties in the case of the expected limits.
}
\label{fig:ExCont}
\end{figure}

The observed and expected number of events for each signal mass point and their
corresponding uncertainties are converted into 95\% confidence level (\CL) upper 
limits on the $\stp \stpb$ production cross section in the \mmplane plane.
These are shown by the colored regions in Fig.~\ref{fig:ExCont} as a function of 
$m(\stp)$ and \dm, where the color scale to the right of the figure gives the 
corresponding upper limit values.
The limits are calculated according to the modified frequentist \CLs 
criterion~\cite{frequentist_limit,Read:2002hq,LHC-HCG}. 
A test statistic is defined as the likelihood ratio between the background-only 
and signal-plus-background hypotheses, and is used to set exclusion limits on 
the top squark pair production. The distributions of the test statistic are
built using simulated experiments, where statistical uncertainties are modeled 
with Poisson distributions, and where all systematic uncertainties are modeled 
with a log-normal distribution. When interpreting the results, it is assumed a 
branching fraction of 100\% for the four-body decay scenario. For the combined 
results of the three years, the largest excess in the data corresponds to 2.5 
standard deviations (local significance) for the $\dm = 10\GeV$ SR.

Using the measured upper limits on the top squark pair cross section and the 
theoretical predictions for the cross section, it is determined the 95\% \CL lower
limits on $m(\stp)$ versus \dm.
The solid black line and thick dotted red line in Fig.~\ref{fig:ExCont} give 
the resulting 95\% \CL observed and expected exclusion contours, respectively, 
on $m(\stp)$ as a function of \dm, obtained from combining the 2016, 2017, and 
2018 data. The corresponding thin black lines in Fig.~\ref{fig:ExCont} represent 
the $\pm$1 standard deviation ($\sigma_{\text{theory}}$) variations in the 
limits due to the theoretical uncertainties in the case of the observed limits. 
The thin dashed red lines give the $\pm$1 and $\pm$2 standard deviation 
($\sigma_{\text{experiment}}$) variations in the case of the expected limits, 
coming from the experimental uncertainties.
The maximum sensitivity is reached for the highest \dm ($\dm \approx m(\PW)$), where 
top squark masses up to 700\GeV are excluded. At the lowest \dm value of 10\GeV
covered by the search, the corresponding value is 480\GeV. The reduced sensitivity at 
lower \dm is explained by the lower transverse momentum spectrum of the decay products, 
as shown in Fig.~\ref{fig:presel}, which results in a loss of acceptance.

The limits of the previous analysis are improved. At low \dm the top squark mass
limit is 60\GeV higher, thus improving the sensitivity at low mass splittings 
beyond simple luminosity scaling, while at high \dm the top squark mass limit is 
extended by 140\GeV.
Compared to the results of a similar analysis by the ATLAS Collaboration for the same 
decay mode and final state~\cite{atlasRun2}, the search presented here has comparable 
limits at intermediate and high \dm values. However, at low \dm, the excluded top squark 
mass is 120\GeV higher than the ATLAS limit. This is attributed to a more inclusive 
preselection criteria, where \PQb tagging is not used, and where the discrimination 
between the signal and the dominating \wjets background is done by a multivariate analysis 
tool, whose performance is further enhanced by a BDT specifically trained for each \dm.

\section{Summary}
\label{s:sum}

The results of a search for the direct pair production of top squarks in single-lepton final 
states are presented within a compressed scenario where $R$ parity is conserved and 
the mass difference $\dm = m(\stp) - m(\lsp)$ between the lightest top squark (\stp) and the lightest 
supersymmetric particle, taken to be the lightest neutralino \PSGczDo, does not 
exceed the \PW boson mass. The considered decay mode of the top squark is the 
prompt four-body decay to $\PQb \mathrm{f} \overline{\mathrm{f}}^{\,\prime} \PSGczDo$,
where the fermions in the final state $\mathrm{f}$ and $\overline{\mathrm{f}}^{\,\prime}$ 
represent a charged lepton and its neutrino for the decay products of one \stp, 
and two quarks for the other top squark. 
The search is based on data collected from proton-proton collisions at $\sqrt{s}=13\TeV$, 
recorded with the CMS detector during the years 2016, 2017, and 2018,  
corresponding to an integrated luminosity of 138\fbinv. 
Events are selected containing a single lepton (electron or muon), at 
least one high-momentum jet, and significant missing transverse momentum.
The analysis is based on a multivariate tool specifically trained for different 
\dm regions, thus adapting the signal selection to the evolution of the kinematical
variables as a function of \mmplane.
The dominant background processes are \wjets, \ttbar, and events with 
nonprompt leptons, which are estimated using control regions in the data.

The observed number of events is consistent with the predicted standard model 
backgrounds in all signal regions. Upper limits are set at the 95\% confidence 
level on the $\stp \stpb$ production cross section as a function of the \stp and
\PSGczDo masses, within the context of a simplified model. Assuming a 100\% 
branching fraction in the four-body decay mode, the search excludes top squark 
masses up to 480 and 700\GeV at $\dm = 10$ and 80\GeV, respectively.
The results summarized in this paper are among the best limits to date on 
the top squark pair production cross section, where the top squark decays via 
the four-body mode, and currently correspond to the most stringent limits for
$\dm < 30\GeV$.

\begin{acknowledgments}
We congratulate our colleagues in the CERN accelerator departments for the excellent performance of the LHC and thank the technical and administrative staffs at CERN and at other CMS institutes for their contributions to the success of the CMS effort. In addition, we gratefully acknowledge the computing centers and personnel of the Worldwide LHC Computing Grid and other centers for delivering so effectively the computing infrastructure essential to our analyses. Finally, we acknowledge the enduring support for the construction and operation of the LHC, the CMS detector, and the supporting computing infrastructure provided by the following funding agencies: BMBWF and FWF (Austria); FNRS and FWO (Belgium); CNPq, CAPES, FAPERJ, FAPERGS, and FAPESP (Brazil); MES and BNSF (Bulgaria); CERN; CAS, MoST, and NSFC (China); MINCIENCIAS (Colombia); MSES and CSF (Croatia); RIF (Cyprus); SENESCYT (Ecuador); MoER, ERC PUT and ERDF (Estonia); Academy of Finland, MEC, and HIP (Finland); CEA and CNRS/IN2P3 (France); BMBF, DFG, and HGF (Germany); GSRI (Greece); NKFIH (Hungary); DAE and DST (India); IPM (Iran); SFI (Ireland); INFN (Italy); MSIP and NRF (Republic of Korea); MES (Latvia); LAS (Lithuania); MOE and UM (Malaysia); BUAP, CINVESTAV, CONACYT, LNS, SEP, and UASLP-FAI (Mexico); MOS (Montenegro); MBIE (New Zealand); PAEC (Pakistan); MES and NSC (Poland); FCT (Portugal); MESTD (Serbia); MCIN/AEI and PCTI (Spain); MOSTR (Sri Lanka); Swiss Funding Agencies (Switzerland); MST (Taipei); MHESI and NSTDA (Thailand); TUBITAK and TENMAK (Turkey); NASU (Ukraine); STFC (United Kingdom); DOE and NSF (USA).

\hyphenation{Rachada-pisek} Individuals have received support from the Marie-Curie program and the European Research Council and Horizon 2020 Grant, contract Nos.\ 675440, 724704, 752730, 758316, 765710, 824093, 884104, and COST Action CA16108 (European Union); the Leventis Foundation; the Alfred P.\ Sloan Foundation; the Alexander von Humboldt Foundation; the Belgian Federal Science Policy Office; the Fonds pour la Formation \`a la Recherche dans l'Industrie et dans l'Agriculture (FRIA-Belgium); the Agentschap voor Innovatie door Wetenschap en Technologie (IWT-Belgium); the F.R.S.-FNRS and FWO (Belgium) under the ``Excellence of Science -- EOS" -- be.h project n.\ 30820817; the Beijing Municipal Science \& Technology Commission, No. Z191100007219010; the Ministry of Education, Youth and Sports (MEYS) of the Czech Republic; the Hellenic Foundation for Research and Innovation (HFRI), Project Number 2288 (Greece); the Deutsche Forschungsgemeinschaft (DFG), under Germany's Excellence Strategy -- EXC 2121 ``Quantum Universe" -- 390833306, and under project number 400140256 - GRK2497; the Hungarian Academy of Sciences, the New National Excellence Program - \'UNKP, the NKFIH research grants K 124845, K 124850, K 128713, K 128786, K 129058, K 131991, K 133046, K 138136, K 143460, K 143477, 2020-2.2.1-ED-2021-00181, and TKP2021-NKTA-64 (Hungary); the Council of Science and Industrial Research, India; the Latvian Council of Science; the Ministry of Education and Science, project no. 2022/WK/14, and the National Science Center, contracts Opus 2021/41/B/ST2/01369 and 2021/43/B/ST2/01552 (Poland); the Funda\c{c}\~ao para a Ci\^encia e a Tecnologia, grant CEECIND/01334/2018 (Portugal); MCIN/AEI/10.13039/501100011033, ERDF ``a way of making Europe", and the Programa Estatal de Fomento de la Investigaci{\'o}n Cient{\'i}fica y T{\'e}cnica de Excelencia Mar\'{\i}a de Maeztu, grant MDM-2017-0765 and Programa Severo Ochoa del Principado de Asturias (Spain); the Chulalongkorn Academic into Its 2nd Century Project Advancement Project, and the National Science, Research and Innovation Fund via the Program Management Unit for Human Resources \& Institutional Development, Research and Innovation, grant B05F650021 (Thailand); the Kavli Foundation; the Nvidia Corporation; the SuperMicro Corporation; the Welch Foundation, contract C-1845; and the Weston Havens Foundation (USA).
\end{acknowledgments}
\bibliography{auto_generated}
\cleardoublepage \appendix\section{The CMS Collaboration \label{app:collab}}\begin{sloppypar}\hyphenpenalty=5000\widowpenalty=500\clubpenalty=5000
\cmsinstitute{Yerevan Physics Institute, Yerevan, Armenia}
{\tolerance=6000
A.~Tumasyan\cmsorcid{0009-0000-0684-6742}
\par}
\cmsinstitute{Institut f\"{u}r Hochenergiephysik, Vienna, Austria}
{\tolerance=6000
W.~Adam\cmsorcid{0000-0001-9099-4341}, J.W.~Andrejkovic, T.~Bergauer\cmsorcid{0000-0002-5786-0293}, S.~Chatterjee\cmsorcid{0000-0003-2660-0349}, K.~Damanakis\cmsorcid{0000-0001-5389-2872}, M.~Dragicevic\cmsorcid{0000-0003-1967-6783}, A.~Escalante~Del~Valle\cmsorcid{0000-0002-9702-6359}, P.S.~Hussain\cmsorcid{0000-0002-4825-5278}, M.~Jeitler\cmsAuthorMark{1}\cmsorcid{0000-0002-5141-9560}, N.~Krammer\cmsorcid{0000-0002-0548-0985}, L.~Lechner\cmsorcid{0000-0002-3065-1141}, D.~Liko\cmsorcid{0000-0002-3380-473X}, I.~Mikulec\cmsorcid{0000-0003-0385-2746}, P.~Paulitsch, F.M.~Pitters, J.~Schieck\cmsAuthorMark{1}\cmsorcid{0000-0002-1058-8093}, R.~Sch\"{o}fbeck\cmsorcid{0000-0002-2332-8784}, D.~Schwarz\cmsorcid{0000-0002-3821-7331}, M.~Sonawane\cmsorcid{0000-0003-0510-7010}, S.~Templ\cmsorcid{0000-0003-3137-5692}, W.~Waltenberger\cmsorcid{0000-0002-6215-7228}, C.-E.~Wulz\cmsAuthorMark{1}\cmsorcid{0000-0001-9226-5812}
% ...
\par}
\vskip\cmsinstskip
\dag:~Deceased\\
$^{1}$Also at TU Wien, Vienna, Austria\\
% ...
\end{sloppypar}
\end{document}